\documentclass{article}

\usepackage{arxiv}

\usepackage[utf8]{inputenc} 
\usepackage[T1]{fontenc}    
\usepackage{hyperref}       
\usepackage{url}            
\usepackage{booktabs}       
\usepackage{amsfonts}       
\usepackage{amsmath}
\usepackage{float}
\usepackage{subcaption}

\usepackage{algorithm2e}
\usepackage{nicefrac}       
\usepackage{microtype}      
\usepackage{lipsum}
\usepackage{graphicx}

\graphicspath{ {./images/} }

\title{Fully Differentiable Ultrasound Simulation Utilizing Ray-Tracing}

\author{
L. River Spencer \\
Department of Aerospace Engineering \& Engineering Mechanics,\\
The University of Texas at Austin,\\
Austin, TX 78712
\And
Reagan A. Cardoza \\
Department of Aerospace Engineering \& Engineering Mechanics,\\
The University of Texas at Austin,\\
Austin, TX 78712
\And
Vijay K. Dubey \\
Walker Department of Mechanical Engineering,\\
The University of Texas at Austin,\\
Austin, TX 78712
\And
Collin E. Haese \\
Walker Department of Mechanical Engineering,\\
The University of Texas at Austin,\\
Austin, TX 78712
\And
Felix Kreidel \\
Clinic for Cardiology,\\
Asklepios Hospital Harburg,\\
Hamburg, Germany
\And
Issam Moussa \\
Carle Heart and Vascular Institute, Carle Health,\\
University of Illinois Urbana-Champaign Carle Illinois College of Medicine,\\
Urbana, IL 61801
\And
Manuel K. Rausch \\
Department of Aerospace Engineering \& Engineering Mechanics,\\
The Oden Institute of Computational Science and Engineering,\\
Department of Biomedical Engineering,\\
Walker Department of Mechanical Engineering,\\
The University of Texas at Austin,\\
Austin, TX 78712
\And
Jan N. Fuhg \\
Department of Aerospace Engineering \& Engineering Mechanics,\\
The Oden Institute of Computational Science and Engineering,\\
The University of Texas at Austin,\\
Austin, TX 78712
}

\begin{document}
\maketitle
\begin{abstract}
Ultrasound imaging applications such as calibration, inverse parameter estimation, and acquisition design require computational models that are physically grounded, efficient, and differentiable with respect to meaningful material and system parameters. While full-wave solvers provide high physical fidelity, they are often prohibitively expensive for iterative optimization, and existing ray-based simulators have largely been used only as forward models. In this work, we present a fully differentiable end-to-end ultrasound simulation framework based on full-path Monte Carlo ray tracing. Building on the UltraRay formulation, the proposed method propagates gradients from image-space objectives back through acoustic transport, beamforming, and post-processing, enabling gradient-based optimization over scene and acquisition parameters.

The framework combines differentiable ray transport in Mitsuba~3/Dr.Jit with a custom differentiable bridge through the ultrasound image-formation pipeline. To obtain stable sensitivities in the Monte Carlo setting, automatic differentiation is applied to fixed sampled ray realizations with respect to continuous parameters. The proposed approach is validated in both forward and inverse settings. Forward examples demonstrate reproduction of expected geometric image features and the ability to represent more complex anatomical structures. In a simulated-reference inverse problem, the method recovers known scene parameters from image-space supervision alone. In a simulation-to-real experiment, it identifies an effective set of parameters that improves agreement between simulated and experimentally acquired B-mode images. Finite-difference comparisons further support the consistency of the computed automatic-differentiation gradients. Overall, this work establishes a practical foundation for differentiable, physics-parameterized ultrasound simulation and optimization.
\end{abstract}


\section{Introduction}
Ultrasound imaging plays a central role in modern clinical practice due to its safety, portability, real-time capability, and relatively low cost \cite{leighton2007ultrasound}. It is widely used across cardiology \cite{villemain2020ultrafast}, obstetrics \cite{woo2002short}, oncology \cite{kadakia2023ultrasound}, and emergency medicine \cite{osterwalder2023point}, where image quality directly affects diagnostic reliability and clinical decision making. The formation of ultrasound images depends on physical and acquisition parameters, including tissue properties, transducer geometry, beamforming strategies, and operating frequency \cite{laugier2024physical}. Understanding and controlling these factors is essential for system design, calibration, and acquisition optimization, motivating the development of computational models that describe the image formation process.
Once such models are available, they also enable the generation of synthetic ultrasound data under controlled conditions. This capability is increasingly important in machine learning workflows, where large, well-annotated clinical datasets are often difficult, costly, or impractical to obtain \cite{erickson2017machine}. Synthetic data can supplement real acquisitions, provide controlled variations of physical parameters, and support training, validation, and benchmarking of data-driven methods, see recent review by Ref. \cite{paproki2024synthetic}. Computational modeling, therefore, serves both as a tool for principled system analysis and as an enabler of modern machine learning applications in ultrasound imaging.
Building on the availability of synthetic data and increasing computational power, data-driven approaches have emerged as an alternative means of modeling ultrasound images. Generative models, including adversarial networks and related neural architectures, have been used to synthesize realistic B-mode images for training and augmentation purposes \cite{peng2019realtime,pigeau2020ultrasound,reynaud2025echoflow}. In addition, learned surrogate and rendering models have been proposed to approximate aspects of the ultrasound image formation process directly from data \cite{zhang2021learning,guo2024ulre}. Once trained, such models can produce visually convincing images at high computational speed, making them attractive for large-scale data generation and real-time applications.
Despite these advantages, purely data-driven models exhibit structural limitations when the objective extends beyond image synthesis. Data-driven architectures are typically trained to approximate mappings from inputs to images and therefore rely on implicit representations learned from finite datasets. As a result, they do not explicitly parameterize physically meaningful quantities such as acoustic impedance, interface roughness, or transducer geometry \cite{luijten2023ultrasound}. This lack of explicit physical structure can limit interpretability and makes it difficult to directly manipulate or optimize acquisition and material parameters within the model. Furthermore, because these methods learn statistical mappings from data, their ability to generalize is inherently tied to the distribution of the training set \cite{goodfellow2016deep}. As a result, they are typically not designed to support system calibration, parameter identification, or inverse problems that require recovering physical properties directly from measured images \cite{arridge2019solving,ongie2020deep}.
Emerging applications in ultrasound imaging increasingly require models that go beyond visual realism and support direct interaction with underlying physical and acquisition parameters. Examples include optimization over material properties and probe settings \cite{clement2000field,canney2008acoustic}, system identification from measured image data \cite{jensen2002ultrasound}, and task-aware acquisition design in which steering angles, frequencies, or array configurations are tuned for a specific diagnostic objective \cite{synnevag2007adaptive}. Addressing such problems naturally leads to optimization formulations in which losses defined on reconstructed images are minimized with respect to physically meaningful quantities \cite{ongie2020deep}. This, in turn, requires computational models that are differentiable end-to-end with respect to material characteristics and acquisition parameters \cite{chen2023differentiable}.
Taken together, these considerations highlight \textit{a fundamental modeling gap}. On one hand, data-driven methods provide speed and visual realism but lack explicit control over physically meaningful parameters. On the other hand, emerging applications demand models that support calibration, parameter identification, and acquisition optimization directly in terms of material and system properties. Bridging this gap requires computational frameworks that are physically parameterized, computationally efficient, interpretable, and amenable to gradient-based optimization with respect to scene and acquisition characteristics.

To address the need for physically grounded modeling, classical ultrasound simulation approaches solve the heterogeneous acoustic wave equation governing pressure propagation in media with spatially varying density and wave speed \cite{virta2014acoustic}. In diagnostic imaging settings, this typically involves discretizing either the linear second-order pressure formulation or an equivalent first-order pressure–velocity system using, for example, finite-difference \cite{dai1995wave,chiavassa2013wave} or finite-element \cite{van2017finite,royer2022efficient} methods. Widely used open-source toolboxes such as k-Wave \cite{treeby2010kwave,treeby2014modelling} and mSOUND \cite{gu2021msound} provide practical implementations of these full-wave models for heterogeneous and absorbing media. Such formulations explicitly account for variable material parameters and enforce transmission conditions across interfaces where density or sound speed may exhibit jump discontinuities. These methods provide high physical fidelity and can capture complex wave phenomena, including diffraction, interference, multiple scattering, reflection, and refraction. However, this physical accuracy comes at a substantial computational and memory cost, particularly in three-dimensional settings or at clinically relevant frequencies that require fine spatial and temporal resolution. Large computational domains, small grid spacings, and long simulation times lead to high memory consumption and long runtimes. Consequently, while full-wave solvers are well-suited for detailed forward analysis, they are often impractical for large-scale parameter sweeps, real-time applications, or iterative inverse problems that require repeated forward evaluations.
To alleviate the computational burden of full-wave simulations, geometric or ray-based approximations have been proposed as efficient alternatives for modeling ultrasound propagation \cite{glassner1989introduction,peddie2019ray}. In these approaches, acoustic energy is modeled as traveling along discrete rays whose trajectories are governed by geometric optics principles. Interactions with tissue interfaces are treated explicitly through reflection and refraction laws derived from acoustic impedance contrasts, while attenuation and scattering effects are incorporated along ray paths \cite{shirley2008realistic}. By replacing the solution of a global wave equation with the tracing of independent propagation paths, ray-based methods substantially reduce computational cost compared to full-wave solvers. This reduction enables simulation over large domains, rapid parameter sweeps, and repeated forward evaluations that are often required in design and optimization settings.
Recent work has advanced ray-based ultrasound simulation from simple ray casting to full path tracing. In particular, UltraRay \cite{duelmer2026ultraray} introduced a Monte Carlo full-path ray tracer for ultrasound built on the Mitsuba 3 rendering framework \cite{jakob2022mitsuba3}. By explicitly modeling round-trip propagation and secondary reflections rather than injecting echoes directly at interaction points, UltraRay produces more physically consistent B-mode images while remaining computationally efficient. Although UltraRay is implemented within a differentiable rendering framework, its focus remains on forward ultrasound image synthesis rather than gradient-based inverse modeling.
In many emerging applications, however, simulation serves not as an end in itself but as a component within a broader optimization framework. Tasks such as transducer calibration, adaptive beamforming, acquisition design, and inverse estimation of tissue properties from ultrasound images require differentiable end-to-end models that link physically meaningful scene and acquisition parameters to task-specific loss functions defined on reconstructed images. This setting necessitates the propagation of gradients from image-space objectives back to parameters such as acoustic impedance contrasts, interface roughness, probe geometry, center frequency, or transmission steering angles.
In contrast, our work develops and validates a fully differentiable end-to-end pipeline that propagates gradients from image-space losses back to physically meaningful material and acquisition parameters and demonstrates gradient-based calibration and inverse recovery in practice. Building on the full-path ray tracing formulation implemented within Mitsuba~3, we construct an end-to-end auto-differentiable ultrasound simulation framework.

We verify the correctness of the forward model through controlled numerical experiments and validate the simulator by comparing simulated B-mode images with experimentally acquired ultrasound data obtained under known imaging conditions. These comparisons demonstrate that the proposed framework captures salient image characteristics observed in real measurements while maintaining computational efficiency. We further validate the differentiable formulation through gradient-based calibration and inverse recovery experiments, showing that physically meaningful parameters can be identified directly from measured B-mode images and that gradients propagate consistently through the full imaging pipeline.
The remainder of the paper is organized as follows. Section~\ref{sec:ultraray} reviews the underlying ray-tracing formulation and introduces our differentiable extensions to both the transport and image-formation pipeline. Section~\ref{sec:autodiff_and_opt} describes the implementation details, including parameterization strategies, micro-batching schemes, and loss design. Section~\ref{sec:results_and_examples} presents experimental evaluations covering forward validation, gradient-based parameter recovery, and acquisition optimization. Finally, Section~\ref{sec:conclusion} analyzes limitations and failure modes and outlines future directions, including the integration of learned components and more advanced tissue models.

\section{Ray-tracing formulation}
\label{sec:ultraray}
In this section, we first review the ultrasound image-formation pipeline and the quantities it produces, namely channel data and B-mode images. We then summarize the full-path Monte Carlo ray-tracing formulation, including emission, boundary interactions, and the estimator used to accumulate element-wise time signals. Finally, we introduce our differentiable extensions, which expose key scene and acquisition parameters to automatic differentiation and enable gradients to propagate from image-space losses back through beamforming and ray transport.
\subsection{Ultrasound Methodology}
In ultrasound imaging, the forward measurement consists of the received pressure time series at each transducer element. Let $e$ index be a receiver element with surface $A_e$. For a given acquisition event, such as plane-wave transmission, the simulator produces channel data $P(e,t)$, which is subsequently mapped to a B-mode image through a standard signal-processing chain including time-gain compensation, filtering, delay-and-sum beamforming, envelope detection, and log compression.
Under a geometric acoustics approximation, we model returning wavefronts as a distribution of incoming pressure rays. The received pressure at element $e$ and time $t$ can be written as an integral of the incident directional pressure over the element surface and incoming directions:
\begin{equation}
P(e,t)
=
\int_{A_e}\int_{\Omega^+}
P_i(\mathbf{x},t,\boldsymbol{\omega}_i)\,
f_d(\boldsymbol{\omega}_i)\,
d\boldsymbol{\omega}_i\, dA ,
\label{eq:pressure_measurement}
\end{equation}
where $\mathbf{x}\in A_e$ denotes a point on the receiver surface, $\boldsymbol{\omega}_i$ is an incoming direction on the visible hemisphere $\Omega^+$, $P_i$ is the incident directional pressure amplitude reaching the receiver, and $f_d$ encodes the element’s receive directivity.
To model the directional pressure field at a surface point  $\mathbf{x}$, we consider the redistribution of incident pressure into outgoing directions via a local scattering operator. We adopt a path-based viewpoint inspired by physically based rendering, treating ultrasound propagation as transport along rays and representing boundary interactions through a local scattering function that maps an incident direction $\boldsymbol{\omega}_i$ to an outgoing direction $\boldsymbol{\omega}_o$.
In rendering, light transport is described by the rendering equation
\begin{equation}
L_o(\mathbf{x},\boldsymbol{\omega}_o)
=
L_e(\mathbf{x},\boldsymbol{\omega}_o)
+
\int_{\Omega^+}
f_s(\mathbf{x},\boldsymbol{\omega}_i,\boldsymbol{\omega}_o)\,
L_i(\mathbf{x},\boldsymbol{\omega}_i)\,
(\mathbf{n}(\mathbf{x})\cdot\boldsymbol{\omega}_i)\,
d\boldsymbol{\omega}_i ,
\label{eq:light_equation}
\end{equation}
where $L_o$ and $L_i$ denote outgoing and incident radiance, $L_e$ is emitted radiance, $f_s$ is the bidirectional scattering distribution function (BSDF), and $\mathbf{n}(\mathbf{x})$ is the surface normal.
For ultrasound, we adopt an analogous formulation by replacing radiance with a directional pressure amplitude evolving over time. The corresponding transport equation becomes
\begin{equation}
P_o(\mathbf{x},t,\boldsymbol{\omega}_o)
=
P_e(\mathbf{x},t,\boldsymbol{\omega}_o)
+
\int_{\Omega^+}
f_s(\mathbf{x},\boldsymbol{\omega}_i,\boldsymbol{\omega}_o)\,
P_i(\mathbf{x},t,\boldsymbol{\omega}_i)\,
(\mathbf{n}(\mathbf{x})\cdot\boldsymbol{\omega}_i)\,
d\boldsymbol{\omega}_i .
\label{eq:pressure_equation}
\end{equation}
where $P_e$ represents the actively transmitted waveform and is nonzero on the transducer aperture.
Direct evaluation of Eq. \eqref{eq:pressure_equation} at every interaction point, receive element, and time sample is computationally intractable due to the high dimensionality of the integral and the large number of required evaluations. We therefore employ a Monte Carlo ray-tracing strategy \cite{li2018differentiable} in which ray paths are stochastically sampled and their weighted contributions accumulated into element-wise time signals. This yields an unbiased Monte Carlo estimator of the received pressure under the adopted geometric scattering model \cite{veach1998robust}.

\subsection{Monte-Carlo Ray Tracing}
To enable efficient ray tracing while maintaining physical realism, we adopt standard
Monte Carlo ray-tracing (MCRT) strategies following \cite{veach1998robust,pharr2023physically}. 
Starting from the directional and surface integrals defining the received pressure in Eq. \eqref{eq:pressure_measurement}, we write
\begin{equation}
P(e,t)
=
\int_{A_e}\int_{\Omega^+}
P_i(\mathbf{x},t,\boldsymbol{\omega})\, f_d(\boldsymbol{\omega})\,
d\boldsymbol{\omega}\, dA .
\label{eq:pressure_integral}
\end{equation}
Let
\begin{equation}
g(\mathbf{x},t,\boldsymbol{\omega})
=
P_i(\mathbf{x},t,\boldsymbol{\omega})\, f_d(\boldsymbol{\omega})
\end{equation}
denote the integrand, and let $D = A_e \times \Omega^+$ be the domain of integration. We introduce a sampling probability density function $p_t(\mathbf{x},\boldsymbol{\omega})$ on $D$ (with respect to the measure $d\boldsymbol{\omega}\, dA$) that is strictly positive wherever $g(\mathbf{x},t,\boldsymbol{\omega})$ is nonzero. Then Eq. \eqref{eq:pressure_integral} can be rewritten as
\begin{equation}
P(e,t)
=
\mathbb{E}_{(\mathbf{x},\boldsymbol{\omega}) \sim p_t}
\left[
\frac{g(\mathbf{x},t,\boldsymbol{\omega})}{p_t(\mathbf{x},\boldsymbol{\omega})}
\right] .
\label{eq:pressure_expectation}
\end{equation}
Approximating this expectation with a Monte Carlo average over $N$ independent samples $(\mathbf{x}_i,\boldsymbol{\omega}_i) \sim p_t$ yields the estimator
\begin{equation}
\hat{P}(e,t)
=
\frac{1}{N}\sum_{i=1}^{N}
\frac{P_i(\mathbf{x}_i,t,\boldsymbol{\omega}_i)\, f_d(\boldsymbol{\omega}_i)}
     {p_t(\mathbf{x}_i,\boldsymbol{\omega}_i)} .
\label{eq:pressure_mc_general}
\end{equation}
In practice, the sampling density is often factorized as
$p_t(\mathbf{x},\boldsymbol{\omega}) = p_A(\mathbf{x})\, p_\Omega(\boldsymbol{\omega}\,|\,\mathbf{x})$, where $p_A$ governs surface sampling and $p_\Omega$ governs directional sampling \cite{pharr2023physically}. The directional component is typically chosen to reflect the transducer directivity so that directions with larger expected contributions are sampled more frequently, reducing estimator variance through importance sampling. If surface sampling is handled deterministically, the estimator simplifies to
\begin{equation}
\hat{P}(e,t)
=
\frac{1}{N}\sum_{i=1}^{N}
\frac{P_i(\mathbf{x}_i,t,\boldsymbol{\omega}_i)\, f_d(\boldsymbol{\omega}_i)}
     {p_t(\boldsymbol{\omega}_i)} ,
\label{eq:pressure_mc}
\end{equation}
where $\boldsymbol{\omega}_i \sim p_t(\boldsymbol{\omega})$. Under the support condition on $p_t$, this estimator is unbiased with respect to Eq. \eqref{eq:pressure_integral}.

Applying the same change-of-measure argument to the directional integral in the rendering-style transport equation (Eq. \eqref{eq:pressure_equation})  yields the Monte Carlo estimator
\begin{equation}
\hat{P}_o(\mathbf{x},\boldsymbol{\omega}_o)
\approx
P_e(\mathbf{x},\boldsymbol{\omega}_o)
+
\frac{1}{N}\sum_{i=1}^{N}
\frac{
f_s(\mathbf{x},\boldsymbol{\omega}_i,\boldsymbol{\omega}_o)\,
P_i(\mathbf{x},\boldsymbol{\omega}_i)\,
\big(\mathbf{n}(\mathbf{x})\cdot \boldsymbol{\omega}_i\big)
}{
p(\boldsymbol{\omega}_i)
},
\label{eq:render_mc}
\end{equation}
where $\boldsymbol{\omega}_i \sim p(\boldsymbol{\omega})$ are i.i.d.\ samples on $\Omega^+$ drawn from a probability density function $p(\boldsymbol{\omega})$. Each sampled direction contributes to the estimator, weighted by the inverse of its sampling probability, and the result is normalized by the number of samples $N$. In practice, $p(\boldsymbol{\omega})$ is often chosen for importance sampling, for example, proportional to the scattering function and/or the cosine term, in order to reduce variance.

\subsection{Algorithmic Implementation of the Ray-Tracing Model}
In the following, we describe the mechanisms applied at each stage of the simulation and summarize the corresponding physics-based models described in Ref. \cite{duelmer2026ultraray}.
\subsubsection{Ray Emission}
The goal of the emission stage is to specify how acoustic energy is initialized at the transducer aperture $\mathcal{A}$, including the spatial distribution of ray origins $\mathbf{x}_0 \in \mathcal{A}$, their initial propagation directions $\boldsymbol{\omega}_0$, transmit delays $\tau_{\mathrm{tx}}(\mathbf{x}_0)$ associated with steering, and element directivity. In our ray-based formulation, each emitted ray is assigned an initial direction $\boldsymbol{\omega}_0$ and a transmit delay $\tau_{\mathrm{tx}}(\mathbf{x}_0)$ determined by the steering angle. Since geometric ray tracing does not model coherent wave interference, these delays serve as time stamps for accumulating contributions into element-wise time signals rather than simulating field superposition.
The transducer is modeled as a discrete array of $N_e$ elements with element pitch $p$, where $p$ denotes the center-to-center spacing between adjacent elements. The pitch determines the spatial sampling of the aperture and thus controls the effective aperture width $W = (N_e - 1)p$, beam steering geometry, and lateral resolution. Changing $p$ alters the distribution of ray emission locations $\mathbf{x}_0$ across the aperture and modifies the geometric delays used in beamforming, directly influencing the spatial focusing characteristics of the simulated image.
To model element directivity, each emitted ray in direction $\boldsymbol{\omega}_0$ from a surface point with outward normal $\mathbf{n}(\mathbf{x}_0)$ is weighted by
\begin{equation}
f_d(\boldsymbol{\omega}_0) = \max\!\bigl(0,\,\mathbf{n}(\mathbf{x}_0)\cdot \boldsymbol{\omega}_0 \bigr),
\label{eq:directivity}
\end{equation}
which favors directions near the element normal. In the Monte Carlo estimator, normalization is handled by the sampling density and the $\tfrac{1}{N}$ factor rather than within $f_d$.

\subsubsection{Ray Traversal}
The goal of the traversal stage is to propagate emitted rays through the scene, determine their interactions with material interfaces, and update their direction, amplitude, and accumulated travel time according to acoustic transport physics. In the following, we refer to rays traced from the transducer into the scene and propagated through successive boundary interactions as \emph{primary rays}. Rays cast from a surface interaction point toward a selected target point (e.g., a receiver element) are referred to as \emph{secondary rays} \cite{duelmer2026ultraray}. 
After emitting a primary ray from an initial state $(\mathbf{x}_0, \boldsymbol{\omega}_0)$, we follow its path until it either escapes the scene without further intersections (and is terminated) or intersects a boundary at some point $\mathbf{x}$. In the latter case, we (i) apply a transducer sampling strategy (described below) to account for paths that can contribute to the received signal, and (ii) evaluate the local acoustic scattering model at the intersection to sample the direction and weight (amplitude) of the next ray segment.

Reflection and transmission at a planar tissue interface are determined by the acoustic impedance contrast between two media with impedances $Z_1$ (incident medium) and $Z_2$ (transmission medium). Let $\boldsymbol{\omega}_i$ denote the incident ray direction and $\mathbf{n}$ the unit normal at the interface pointing into medium 2. The acoustic impedance ratio is defined as
$
\eta = \frac{Z_1}{Z_2}.
$
Under geometric acoustics, the reflected direction follows specular reflection,
\begin{equation}
\boldsymbol{\omega}_r
=
\boldsymbol{\omega}_i
+
2(\mathbf{n}\cdot(-\boldsymbol{\omega}_i))\,\mathbf{n},
\end{equation}
while the transmitted direction $\boldsymbol{\omega}_t$ is obtained from Snell’s law. The reflection amplitude is determined by the acoustic Fresnel coefficient
\begin{equation}
A_r =
\frac{Z_1 \cos\theta_r - Z_2 \cos\theta_t}
     {Z_1 \cos\theta_r + Z_2 \cos\theta_t},
\end{equation}
where $\theta_r$ and $\theta_t$ denote the reflection and transmission angles. The squared magnitude $(A_r)^2$ specifies the fraction of incident intensity reflected at the interface and therefore defines the probabilistic energy split between reflected and transmitted rays in the Monte Carlo scheme.
Following Ref.~\cite{duelmer2026ultraray}, we model surface roughness using a microfacet distribution, allowing a continuous transition between specular and diffuse scattering. Instead of a single planar interface with uniform normal $\mathbf{n}$, the boundary is represented as a distribution of microfacet normals whose normal distribution function is given by
\begin{equation}
    D(\mathbf{h}) =
    \frac{\alpha^2}{\pi\left((\mathbf{n}\cdot \mathbf{h})^2(\alpha^2 - 1) + 1\right)^2},
    \label{eq:GGX_PDF}
\end{equation}
where $\alpha$ is the roughness parameter, $\mathbf{h} = (\boldsymbol{\omega}_i + \boldsymbol{\omega}_o)/\|\boldsymbol{\omega}_i + \boldsymbol{\omega}_o\|$ is the half-vector, and $\mathbf{n}$ is the macroscopic surface normal. Smaller $\alpha$ values concentrate microfacet orientations near $\mathbf{n}$ and produce predominantly specular reflections, whereas larger $\alpha$ values yield broader orientation distributions and more diffuse scattering.
Introducing a microfacet-based roughness description allows the boundary model to interpolate smoothly between mirror-like and diffuse scattering regimes, yielding a more realistic treatment of acoustic reflections at tissue interfaces. In the Monte Carlo estimator of Eq.~\eqref{eq:render_mc}, this microfacet formulation modifies both the scattering operator $f_s$ and the corresponding sampling distribution $p(\boldsymbol{\omega}_i)$, thereby controlling the angular redistribution and weighting of reflected rays. For computational tractability, ray propagation is limited by enforcing a maximum number of successive interactions per path.

\subsubsection{Transducer Sampling}
To improve the probability that simulated transport paths produce a nonzero receiver contribution, we use a receiver-guided sampling strategy. After each surface interaction, we generate a receiver-directed secondary ray that is aimed at a point on the transducer. Concretely, we first choose a transducer element uniformly at random and set the aim point to that element's center, we then trace the corresponding secondary ray through the scene, accumulating pressure attenuation terms in the same way as for primary rays. If the ray intersects another boundary, the BSDF at that interaction is evaluated and the path throughput is updated accordingly. Because this sampling explicitly enforces a direction that connects the interaction point to the receiver, we include the probability of selecting and continuing along this receiver-directed direction in the path weight. Operationally, this follows the same interaction/throughput bookkeeping as the primary transport model, but with the direction constrained to the sampled receiver point.

We also model the receive directivity of each transducer element by applying a direction weight \(f_d\) that depends on the incident direction \(\omega_i\) and the element normal \(\mathbf{n}\) (mirroring the transmit-side treatment). Using a main-lobe angle \(\alpha_m\) and a cutoff angle \(\alpha_c\), we define 

\begin{equation}
    f_d(\omega_i) =
    \begin{cases}
        0, & |\alpha| > \alpha_c, \\
        \frac{\alpha_c - |\alpha|}{\alpha_c - \alpha_m}, & \alpha_m < |\alpha| \leq \alpha_c, \\
        1, & |\alpha| \leq \alpha_m,
    \end{cases}
    \label{eq:cutoff}
\end{equation}

where \(\alpha\) is the angle between \(\omega_i\) and \(\mathbf{n}\), computed as \(\alpha = \arccos(\mathbf{n}\cdot \omega_i)\). This factor can be incorporated directly into Eq.~\eqref{eq:pressure_mc}.

\subsubsection{Phase Calculation and Signal Processing}
When a simulated path intersects the transducer, the remaining pressure amplitude carried by that ray is deposited into the receive time series of the corresponding transducer element at the appropriate time-of-flight. To model the finite bandwidth of the transmit/receive chain and to introduce an oscillatory phase consistent with the center frequency, we filter this impulse-like contribution with a band-limited pulse. In our implementation, we use a sinusoid tapered by a Gaussian envelope, which approximates the axial pulse response (i.e., the axial point-spread function). The pulse shape is

\begin{equation}
    s(t) = \sin(2\pi f_c t)\,\exp\!\left(-\frac{t^2}{\sigma}\right),
    \label{eq:axial_sig}
\end{equation}

where \(f_c\) denotes the center frequency of the transducer and \(\sigma\) controls the envelope width. We choose \(\sigma\) to match the desired number of emitted wave cycles, thereby setting the effective bandwidth and axial resolution.

After all ray contributions have been accumulated and filtered across the element-wise time series, the ray-tracing stage produces channel data for the full aperture. This channel data is then passed through a conventional ultrasound processing chain: plane-wave delay-and-sum beamforming, demodulation to baseband, and logarithmic compression. Finally, we apply dynamic-range limiting and convert the results to grayscale to form the B-mode image.

\section{Automatic Differentiation and Optimization}
\label{sec:autodiff_and_opt}

Our goal is to enable gradient-based calibration and inverse design over physically meaningful parameters of the ultrasound pipeline. Concretely, we optimize a parameter vector \(\theta\) (e.g., interface roughness, acoustic impedance, probe pitch, center frequency) to minimize a scalar loss defined on the final B-mode image.

\subsection{End-to-end objective}

Let \(\hat{\mathbf{P}}(\theta)\in\mathbf{R}^{N_e\times N_t}\)  denote the simulated channel data (element-by-time pressure signals) produced by the full-path MCRT ray tracer, and let \(\mathcal{B}(\cdot)\) denote the beamforming and post-processing operator that maps channel data to a B-mode image (delay-and-sum beamforming, demodulation/envelope detection, log compression, and dynamic-range clipping), consistent with Ultra Ray's imaging pipeline.

Given a reference image \(\mathbf{I}_{\text{ref}}\) (synthetic or measured), we minimize

\begin{equation}
\min_{\theta}\ \mathcal{L}(\theta)
\;=\;
\ell\!\left(\mathbf{I}(\theta),\,\mathbf{I}_{\text{ref}}\right),
\qquad
\mathbf{I}(\theta) \;=\; \mathcal{B}\!\left(\hat{\mathbf{P}}(\theta)\right),
\label{eq:opt_objective}
\end{equation}
where \(\ell(\cdot,\cdot)\) is an image-space loss.

\subsection{Differentiating through Monte Carlo ray transport}

The ray tracer produces \(\hat{\mathbf{P}}(\theta)\) via a Monte Carlo estimator (Eq.~\eqref{eq:pressure_mc}) built from sampled paths, scattering decisions, and continuous weights (BSDF terms, directivity, geometric factors).The Mitsuba 3 / Dr.Jit backend supports reverse-mode automatic differentiation of such estimators with respect to continuous scene parameters that influence intersection geometry and the BSDF (e.g., GGX roughness \(\alpha\) and impedance \(Z\) via Fresnel terms), as long as the random numbers used to sample paths are treated as fixed during differentiation. This corresponds to a standard \emph{reparameterized} Monte Carlo gradient estimator: we draw a set of random samples \(\xi\) (ray origins/directions, BSDF sampling decisions, etc.), hold \(\xi\) constant for the backwards pass, and differentiate the resulting deterministic computation graph with respect to \(\theta\).

Denoting a single acquisition realization as
\(
\hat{\mathbf{P}}(\theta;\xi)
\),
the gradient of Eq.~\eqref{eq:opt_objective} follows the chain rule
\begin{equation}
\nabla_{\theta}\mathcal{L}
\;=\;
\left(\frac{\partial \hat{\mathbf{P}}}{\partial \theta}\right)^{\!\top}
\,
\frac{\partial \mathcal{L}}{\partial \hat{\mathbf{P}}},
\label{eq:chain_rule}
\end{equation}

where \(\tfrac{\partial \hat{\mathbf{P}}}{\partial \theta}\) is obtained by Dr.Jit reverse-mode AD through the ray tracing kernels, and \(\tfrac{\partial \mathcal{L}}{\partial \hat{\mathbf{P}}}\) is the adjoint of the beamforming / post-processing chain (detailed in the next section).

In practice, we reduce gradient noise by using \emph{common random numbers} across optimization iterations: the sampler seeds and ray sampling sequences are kept fixed while \(\theta\) changes. This decreases the variances of \(\nabla_{\theta}\mathcal{L}\) without biasing the gradient of the sampled objective (it effectively makes the optimization follow the gradient of a fixed Monte Carlo realization). Accordingly, the backward pass does not propagate derivatives through the stochastic sampling mechanism itself; instead, for fixed random samples, it differentiates the deterministic transport, beamforming, and image-processing operations with respect to the continuous optimization parameters.

\subsection{Differentiating through beamforming and post-processing}
\label{sec:bf_diff}

The beamforming operator \(\mathcal{B}\) is not natively expressed in Dr.Jit: it involves a signal-processing chain (implemented using Ultraspy). We therefore treat beamforming as an \emph{external operator} and provide its vector-Jacobian product (VJP) to Dr.Jit so that gradients can flow from image-space losses back to ray-traced channel data.

A key observation is that most of the pipeline is either linear or composed of simple pointwise nonlinearities: (i) axial convolution (Eq.~\eqref{eq:axial_sig}) is linear, (ii) delay-and-sum (DAS) beamforming is linear in the channel data, (iii) envelope detection and log compression are pointwise nonlinear operations.

\paragraph{Delay-and-sum as a linear operator.}
Let \(\mathbf{I}_{\text{rf}}(x,z)\) denote the beamformed RF image before demodulation/log compression at pixel \((x,z)\). In DAS plane-wave imaging, a standard discrete form is 

\begin{equation}
\mathbf{I}_{\text{rf}}(x,z)
=
\sum_{e=1}^{N_e}
w_e(x,z)\,
\hat{\mathbf{P}}\!\left(e,\; \tau_e(x,z)\right),
\label{eq:das}
\end{equation}

where \(w_e\) are apodization/directivity weights and \(\tau_e(x,z)\) is the sample time corresponding to the two-way travel time for the chosen transmit event and receive element geometry. Since \(\tau_e(x,z)\) is generally not an integer sample index, \(\hat{\mathbf{P}}(e,\tau_e)\) is evaluated via linear interpolation, which preserves linearity in \(\hat{\mathbf{P}}\).

Because Eq.~\eqref{eq:das} is linear in \(\hat{\mathbf{P}}\), its adjoint (backpropagation) is a \emph{backprojection} that scatters image gradients back into the channel-time grid using the same interpolation weights:

\begin{equation}
\frac{\partial \mathcal{L}}{\partial \hat{\mathbf{P}}(e,t)}
=
\sum_{x,z}
w_e(x,z)\;
\frac{\partial \mathcal{L}}{\partial \mathbf{I}_{\text{rf}}(x,z)}
\;
\frac{\partial \hat{\mathbf{P}}(e,\tau_e(x,z))}{\partial \hat{\mathbf{P}}(e,t)}.
\label{eq:das_adjoint}
\end{equation}

The last factor is nonzero only for the neighboring samples used in the linear interpolation at \(\tau_e(x,z)\), resulting in an efficient scatter-add implementation.

\paragraph{Custom bridge between Dr.Jit and beamforming.}
We implement the mapping \(\hat{\mathbf{P}} \mapsto \mathbf{I}=\mathcal{B}(\hat{\mathbf{P}})\) as a custom differentiable operator: the forward pass converts Dr.Jit arrays to the processing backend, runs \(\mathcal{B}\), and returns \(\mathbf{I}\); the backward pass takes the upstream image gradient \(\tfrac{\partial \mathcal{L}}{\partial \mathbf{I}}\), computes \(\tfrac{\partial \mathcal{L}}{\partial \hat{\mathbf{P}}}\) via the adjoint of \(\mathcal{B}\) (Eq.~\eqref{eq:das_adjoint} and Eq.~\eqref{eq:log_env_grad}), and \emph{seeds} Dr.Jit’s reverse-mode AD at the channel data tensor \(\hat{\mathbf{P}}\). Dr.Jit then propagates this adjoint through the ray tracer to obtain \(\nabla_{\theta}\mathcal{L}\). This design allows end-to-end differentiation while keeping the high-performance ray transport in Mitsuba and the signal processing in specialized kernels \cite{duelmer2026ultraray}.

\subsection{Parameterization and constraints}
\label{sec:param_constraints}

We optimize a set of physically meaningful parameters (e.g., GGX roughness, acoustic impedance, transducer pitch, and center frequency) that must remain within prescribed admissible ranges. Rather than optimizing these quantities directly, we maintain unconstrained optimization variables \(\tilde{\theta}\in\mathbb{R}^d\) and map them into normalized variables \(\theta_n\in[0,1]^d\), which are then linearly scaled into physical parameter ranges.

\paragraph{Normalization to \([0,1]\) and scaling to physical ranges.} For each parameter \(j\), we define a minimum and maximum admissible value \((\theta^{\min}_j,\theta^{\max}_j)\). We map the optimizer state \(\tilde{\theta}_j\) into a normalized variable by hard clipping:
\begin{equation}
\theta_{n,j} = \mathrm{clip}\!\left(\tilde{\theta}_j,\,0,\,1\right),
\label{eq:clip01}
\end{equation}
and then scale to the physical value used in the simulator by
\begin{equation}
\theta_j
=
\theta^{\min}_j
+
\theta_{n,j}\left(\theta^{\max}_j-\theta^{\min}_j\right).
\label{eq:scale_phys}
\end{equation}

This parameterization makes all optimization variables dimensionless and comparably scaled, which simplifies learning-rate selection across heterogeneous parameters (e.g., roughness versus frequency).

\paragraph{Hard clipping.} We enforce feasibility by clamping \(\theta_{n,j}\) to \([0,1]\) (Eq.~\eqref{eq:clip01}) at every iteration before running the forward simulation. When gradients are computed, the backward pass uses the local derivative of the clip operator:
\begin{equation}
\frac{\partial \theta_{n,j}}{\partial \tilde{\theta}_j}
=
\begin{cases}
1, & 0 < \tilde{\theta}_j < 1,\\
0, & \tilde{\theta}_j \le 0 \ \text{or}\ \tilde{\theta}_j \ge 1,
\end{cases}
\label{eq:clip_grad}
\end{equation}
so parameters that saturate at a bound receive zero gradient until the optimizer updates move them back into the interior. In practice, this simple constraint handling proved sufficient for the calibration and inverse-design experiments considered in this work.

\paragraph{Examples.} Using this scheme, the simulator parameters are obtained as
\(\rho = \rho^{\min} + \theta_{n,\rho}(\rho^{\max}-\rho^{\min})\) for surface roughness,
\(Z = Z^{\min} + \theta_{n,Z}(Z^{\max}-Z^{\min})\) for acoustic impedance, and
\(R = R^{\min} + \theta_{n,R}(R^{\max}-R^{\min})\) for cylinder radius.

\subsection{Micro-batched acquisition and gradient estimation}
\label{sec:microbatch}

Directly differentiating a full plane-wave acquisition with many steering angles and a large number of emitted rays per element can exceed memory due to AD tape storage and intermediate buffers. To make optimization tractable, we use micro-batching across both (i) transmit angles and (ii) subsets of emitted rays/elements. Each micro-batch produces a partial channel-data tensor \(\hat{\mathbf{P}}_b(\theta)\) and an associated image and loss. We accumulate losses across micro-batches:
\begin{equation}
\mathcal{L}(\theta)
\approx
\sum_{b=1}^{B} \mathcal{L}_b(\theta),
\qquad
\mathcal{L}_b(\theta)=
\ell\!\left(\mathcal{B}\!\left(\hat{\mathbf{P}}_b(\theta)\right),\,\mathbf{I}_{\text{ref},b}\right),
\label{eq:microbatch_loss}
\end{equation}

and backpropagate each \(\mathcal{L}_b\) independently, accumulating gradients in \(\nabla_{\theta}\mathcal{L}\). This is equivalent to full-batch optimization when the partitioning is exact (deterministic split of rays/angles) and provides a low-memory approximation otherwise (stochastic split), analogous to minibatch stochastic gradient descent.

\subsection{Loss function}
\label{sec:losses}

All optimization experiments in this work use a weighted combination of \(\ell_1\) and \(\ell_2\) image losses defined directly on the final B-mode images. Let \(\mathbf{I}(\theta)=\mathcal{B}(\hat{\mathbf{P}}(\theta))\) denote the simulated B-mode image produced by the end-to-end pipeline for parameters \(\theta\), and let \(\mathbf{I}_{\mathrm{ref}}\) denote the reference B-mode image. The optimization objective is defined as

\begin{equation}
\mathcal{L}(\theta)
=
\lambda_{1}\left\|\mathbf{I}(\theta)-\mathbf{I}_{\mathrm{ref}}\right\|_{1}
+
\lambda_{2}\left\|\mathbf{I}(\theta)-\mathbf{I}_{\mathrm{ref}}\right\|_{2}^{2},
\label{eq:l1_l2_loss}
\end{equation}

which may be written componentwise as

\begin{equation}
\mathcal{L}(\theta)
=
\lambda_{1}\sum_{u}\left|\mathbf{I}_{u}(\theta)-\mathbf{I}_{\mathrm{ref},u}\right|
+
\lambda_{2}\sum_{u}\left(\mathbf{I}_{u}(\theta)-\mathbf{I}_{\mathrm{ref},u}\right)^{2},
\label{eq:l1_l2_loss_expanded}
\end{equation}

where \(u\) indexes image pixels. The \(\ell_1\) term promotes robustness to localized intensity mismatches, while the \(\ell_2\) term penalizes larger deviations more strongly. In practice, we normalize this objective by the number of pixels to make the loss invariant to image resolution:

\begin{equation}
\mathcal{L}_{\mathrm{img}}(\theta)
=
\frac{1}{N_{\mathrm{pix}}}
\left[
\lambda_{1}\sum_{u}\left|\mathbf{I}_{u}(\theta)-\mathbf{I}_{\mathrm{ref},u}\right|
+
\lambda_{2}\sum_{u}\left(\mathbf{I}_{u}(\theta)-\mathbf{I}_{\mathrm{ref},u}\right)^{2}
\right].
\label{eq:weighted_l1_l2_loss}
\end{equation}

Gradients are obtained by backpropagating through the image-formation chain:
\begin{equation}
\frac{\partial \mathcal{L}}{\partial \hat{\mathbf{P}}}
=
\left(\frac{\partial \mathbf{I}}{\partial \hat{\mathbf{P}}}\right)^{\!\top}
\frac{\partial \mathcal{L}}{\partial \mathbf{I}},
\qquad
\frac{\partial \mathcal{L}}{\partial \mathbf{I}}
=
2\left(\mathbf{I}(\theta)-\mathbf{I}_{\mathrm{ref}}\right),
\label{eq:l2_backprop}
\end{equation}
after which Dr.Jit propagates the seeded adjoint through the ray-tracing computation to yield \(\nabla_{\theta}\mathcal{L}\) (Sec.~\ref{sec:bf_diff}).

\subsection{Optimization procedure}

We optimize Eq.~\eqref{eq:opt_objective} using Adam with parameter groups to allow different learning rates for material (roughness/impedance) and acquisition (pitch/frequency/angles) parameters. Let \(\theta_k\) be parameters at iteration \(k\). Adam updates are
\begin{align}
m_k &= \beta_1 m_{k-1} + (1-\beta_1)\nabla_{\theta}\mathcal{L}(\theta_k),
\quad
v_k = \beta_2 v_{k-1} + (1-\beta_2)\nabla_{\theta}\mathcal{L}(\theta_k)^2,\\
\theta_{k+1}
&=
\theta_k
-
\eta
\frac{\hat{m}_k}{\sqrt{\hat{v}_k}+\epsilon},
\label{eq:adam}
\end{align}
with standard bias corrections \(\hat{m}_k,\hat{v}_k\).

We further apply gradient clipping for stability and keep Monte Carlo seeds fixed (common random numbers) unless explicitly stated (Sec.~\ref{sec:bf_diff}).

Algorithm~\ref{alg:opt} summarizes one optimization iteration.

\begin{algorithm}[t]
\caption{One optimization iteration (micro-batched)}
\label{alg:opt}
\begin{enumerate}
\item Select micro-batch \(b\): subset of steering angles and/or rays/elements.
\item Ray trace with Mitsuba/Dr.Jit to obtain channel data \(\hat{\mathbf{P}}_b(\theta)\).
\item Beamform and post-process to \(\mathbf{I}_b=\mathcal{B}(\hat{\mathbf{P}}_b)\); compute loss \(\mathcal{L}_b=\ell(\mathbf{I}_b,\mathbf{I}_{\text{ref},b})\).
\item Backward differentiate through \(\mathcal{B}\) to compute \(\partial \mathcal{L}_b/\partial \hat{\mathbf{P}}_b\); seed Dr.Jit adjoint at \(\hat{\mathbf{P}}_b\).
\item Dr.Jit reverse-mode AD backpropagates through ray tracing to obtain \(\nabla_{\theta}\mathcal{L}_b\).
\item Accumulate gradients over micro-batches and update \(\theta\) using Adam.
\end{enumerate}
\end{algorithm}

\section{Results and Examples}
\label{sec:results_and_examples}

This section evaluates the proposed differentiable ultrasound simulation framework in both forward and inverse settings. The results are organized to assess three central capabilities of the method: first, its ability to generate physically plausible B-mode images from prescribed scene configurations; second, its ability to propagate meaningful gradients through the full imaging pipeline; and third, its ability to support gradient-based recovery of physically interpretable parameters from image-space supervision. Together, these experiments are intended to demonstrate that the framework functions not only as a forward simulator, but also as a practical tool for calibration and inverse analysis.

We begin with forward examples that examine whether the ray-tracing formulation reproduces expected geometric image features in simple and more complex scenes. We then consider two optimization problems based on the hollow-cylinder test case. The first is a simulated-reference experiment, which provides a controlled setting for evaluating parameter recovery and gradient consistency when the target image is generated by the same model. The second is a simulation-to-real experiment, in which the optimized simulation is matched to an experimentally acquired ultrasound image in order to assess the practical applicability of the framework under model mismatch and partial ground-truth uncertainty.

\subsection{Forward Problems}
\label{sec:forward_problems}

\begin{figure}[hbtp]
    \centering
    \begin{subfigure}[t]{0.48\linewidth}
        \centering
        \begin{minipage}[c][0.28\textheight][c]{\linewidth}
            \centering
            \includegraphics[width=\linewidth,height=0.26\textheight,keepaspectratio]{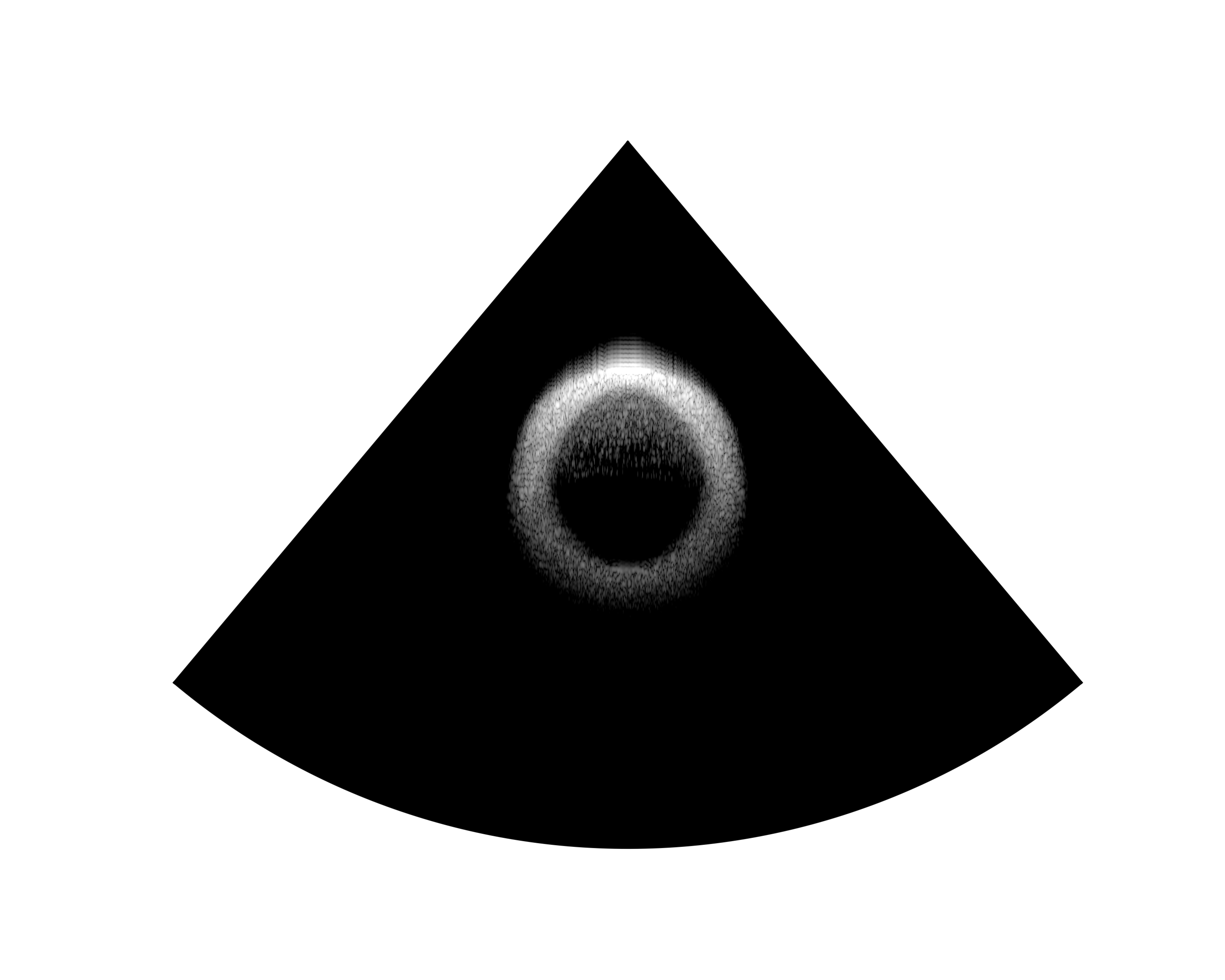}
        \end{minipage}
        \caption{Full cylinder B-mode image.}
        \label{fig:cylinder_cone}
    \end{subfigure}\hfill
    \begin{subfigure}[t]{0.48\linewidth}
        \centering
        \begin{minipage}[c][0.28\textheight][c]{\linewidth}
            \centering
            \includegraphics[width=\linewidth,height=0.26\textheight,keepaspectratio]{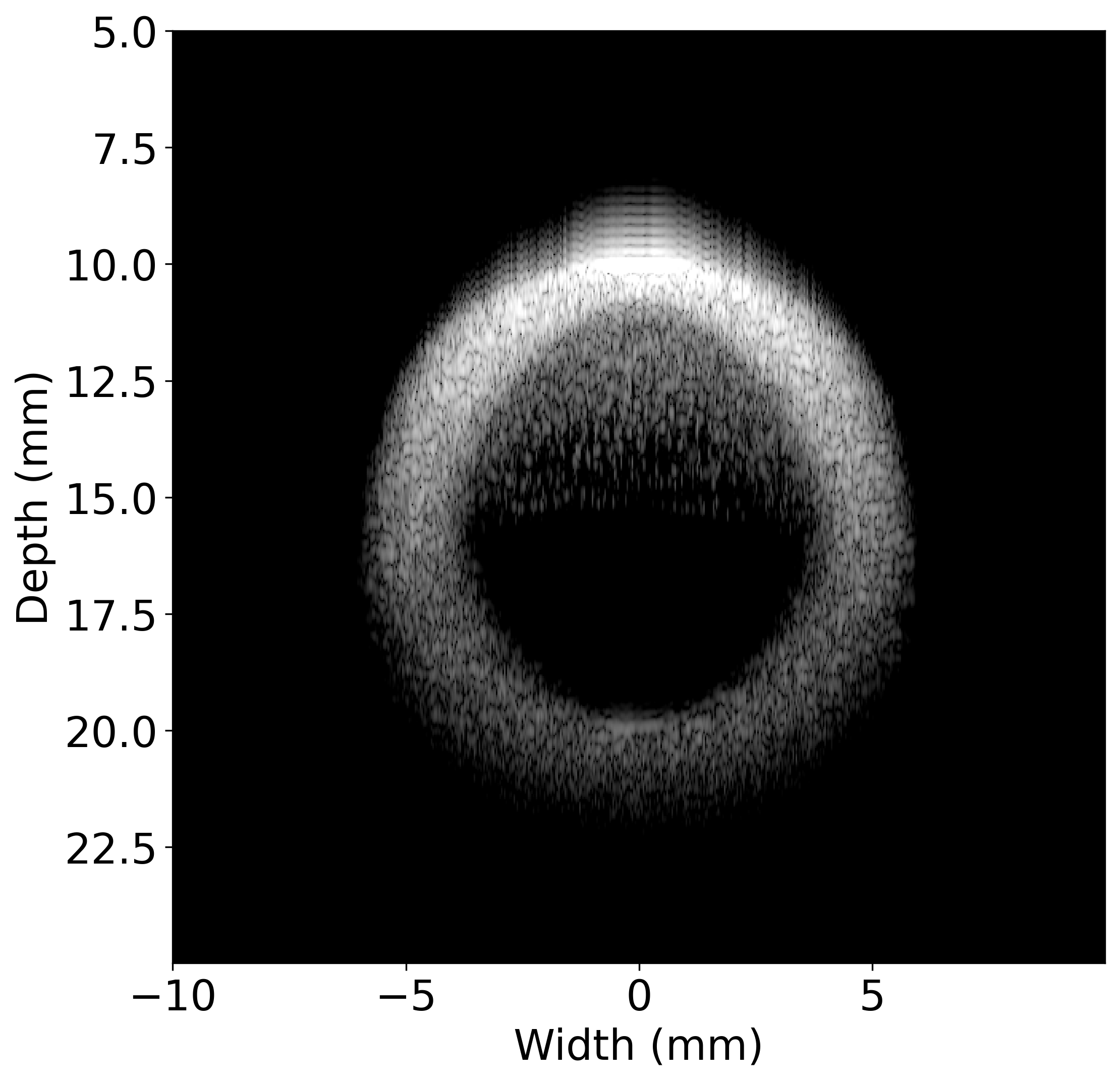}
        \end{minipage}
        \caption{Zoomed-in B-mode image.}
        \label{fig:cylinder_no_cone}
    \end{subfigure}
    \caption{B-mode images for the hollow cylinder test case. \textbf{(a)} The full image shows the overall simulated acquisition geometry. \textbf{(b)} The zoomed-in view highlights the dominant scattering features associated with the cylindrical interface. The observed image structure is consistent with the prescribed cylinder location and radius, supporting the ability of the forward model to reproduce the expected geometric response in a simple canonical configuration.}
    \label{fig:cylinder_forward}
\end{figure}

To assess the accuracy of the forward simulation, we consider two test cases. First, we render a hollow cylinder to confirm that the dimensions measured from the resulting B-mode image agree with the prescribed analytical geometry. The cylinder has a radius of 5 mm and is located 15 mm from the transducer. Figure~\ref{fig:cylinder_cone} shows the full B-mode image, while Figure~\ref{fig:cylinder_no_cone} shows a zoomed-in view.

\begin{figure}[htbp]
    \centering

    \begin{subfigure}[t]{0.48\linewidth}
        \centering
        \includegraphics[width=\linewidth]{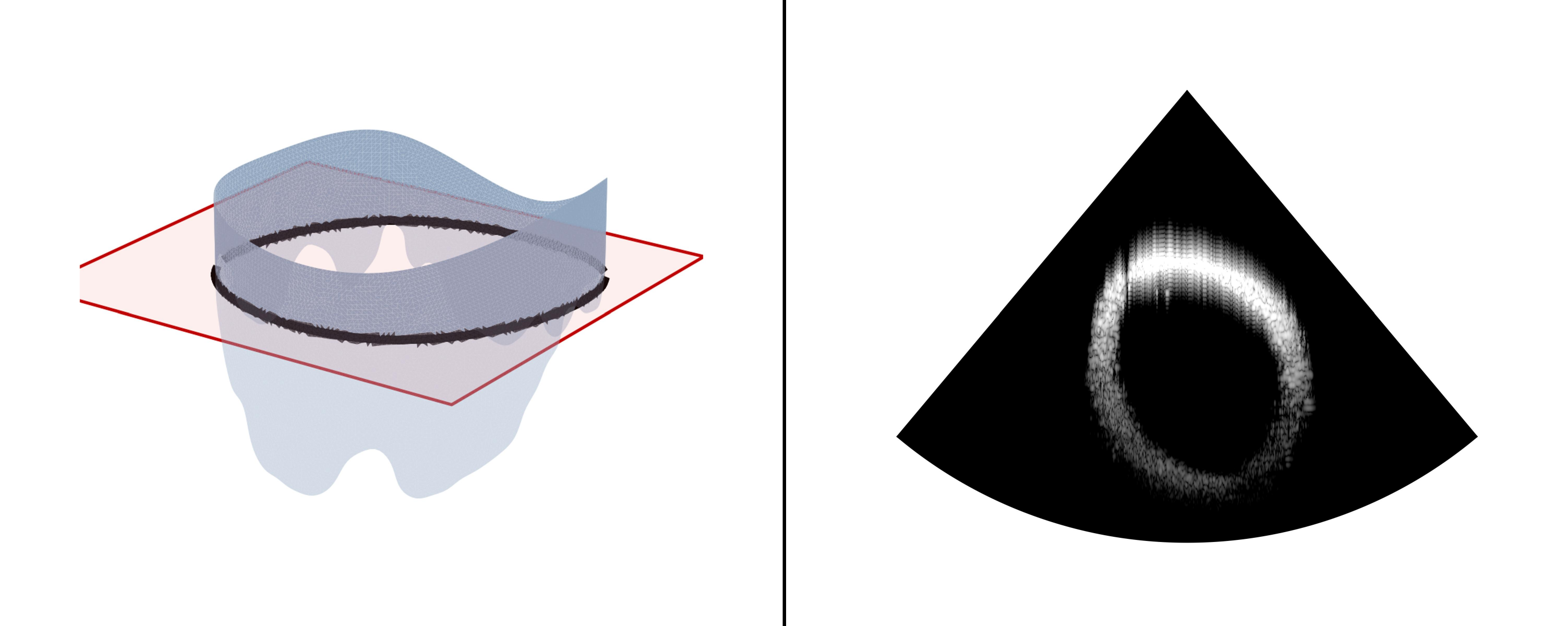}
        \caption{Top down view, $t = 0$ (fully open).}
        \label{fig:top_t0}
    \end{subfigure}\hfill
    \begin{subfigure}[t]{0.48\linewidth}
        \centering
        \includegraphics[width=\linewidth]{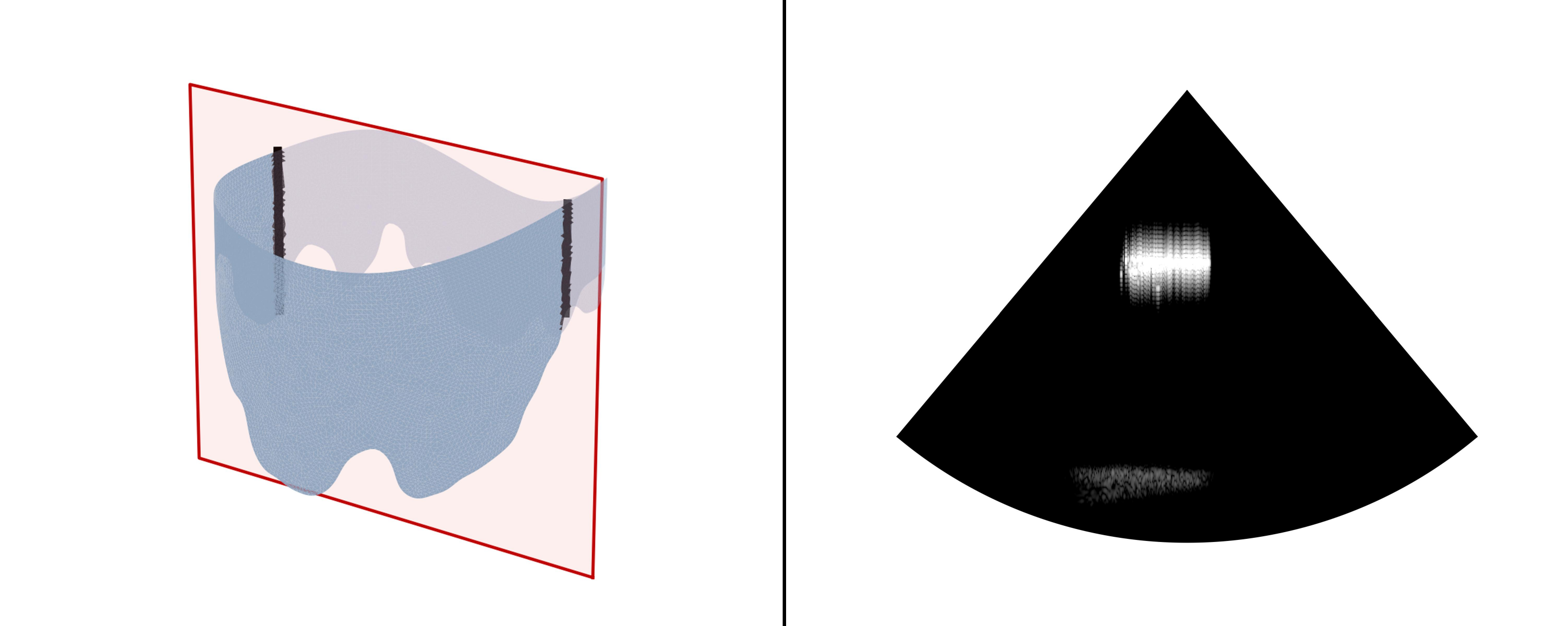}
        \caption{Side view, $t = 0$ (fully open).}
        \label{fig:side_t0}
    \end{subfigure}

    \vspace{0.5em}

    \begin{subfigure}[t]{0.48\linewidth}
        \centering
        \includegraphics[width=\linewidth]{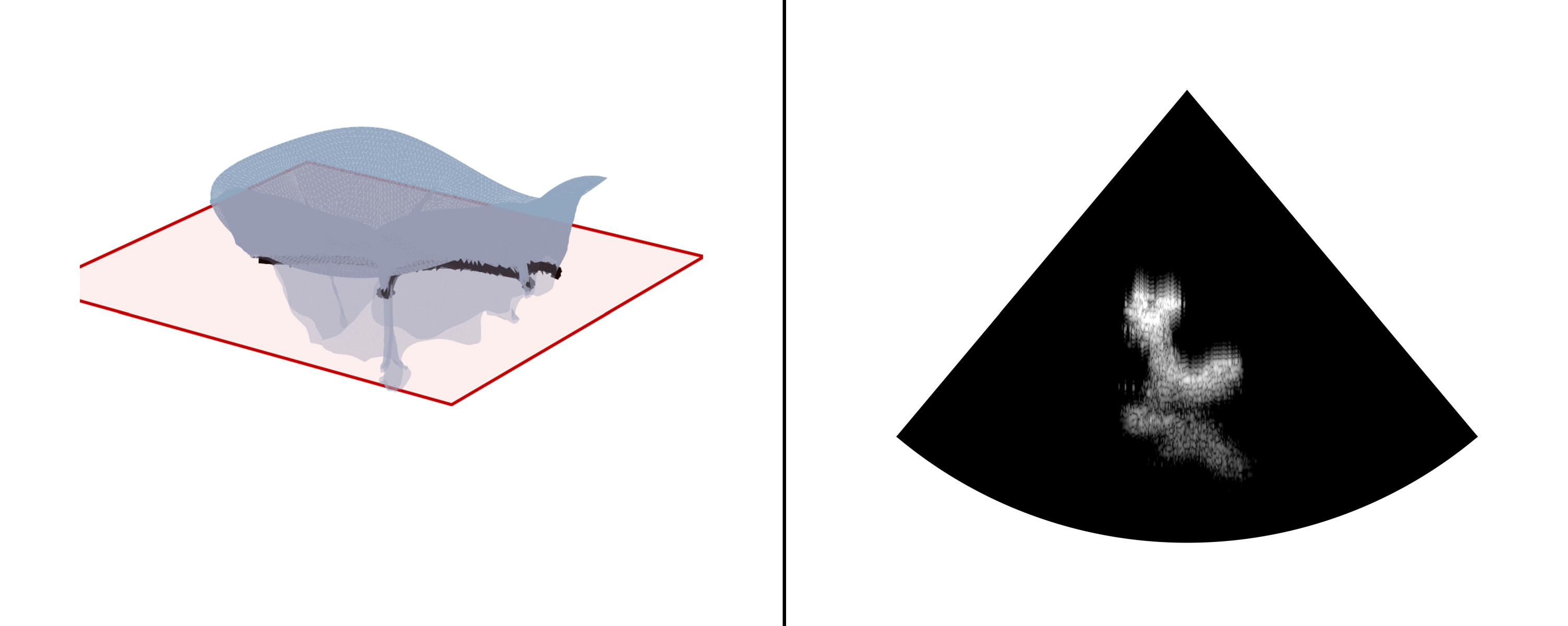}
        \caption{Top down view, $t = 0.5$ (partly closed).}
        \label{fig:top_t05}
    \end{subfigure}\hfill
    \begin{subfigure}[t]{0.48\linewidth}
        \centering
        \includegraphics[width=\linewidth]{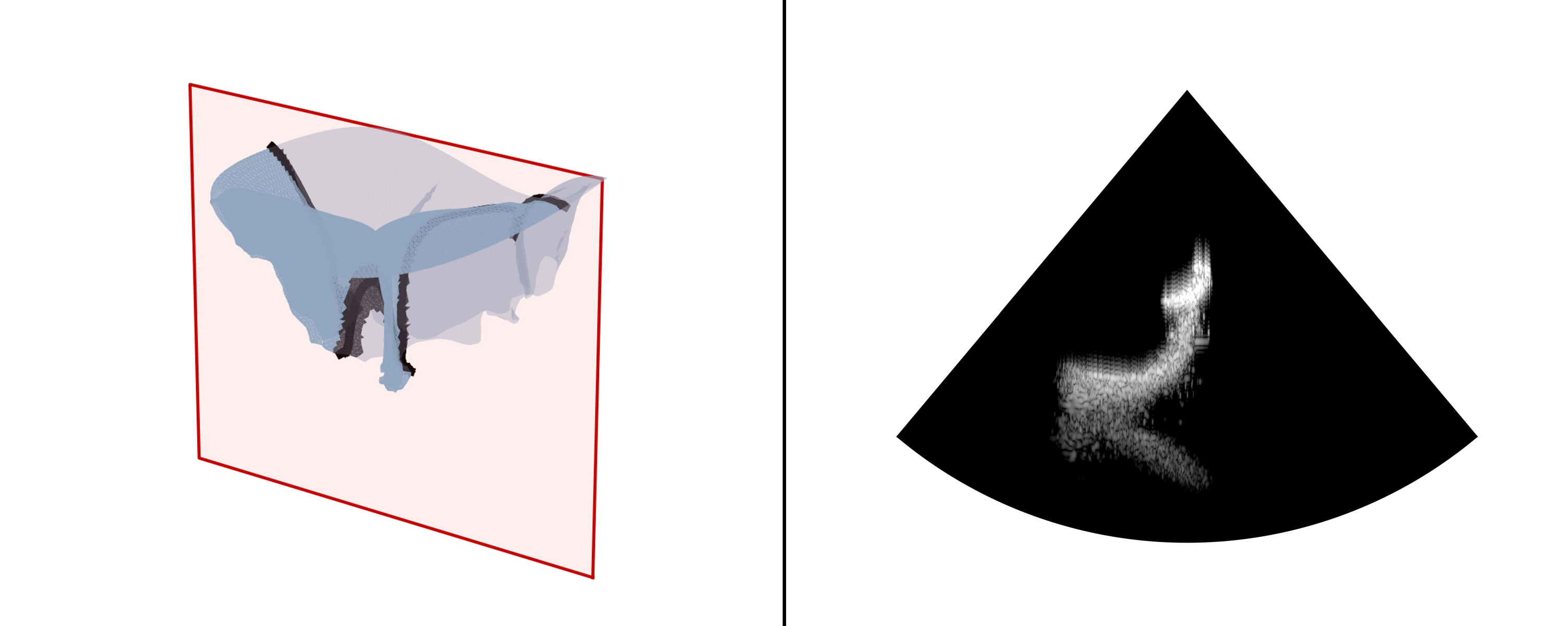}
        \caption{Side view, $t = 0.5$ (partly closed).}
        \label{fig:side_t05}
    \end{subfigure}

    \vspace{0.5em}

    \begin{subfigure}[t]{0.48\linewidth}
        \centering
        \includegraphics[width=\linewidth]{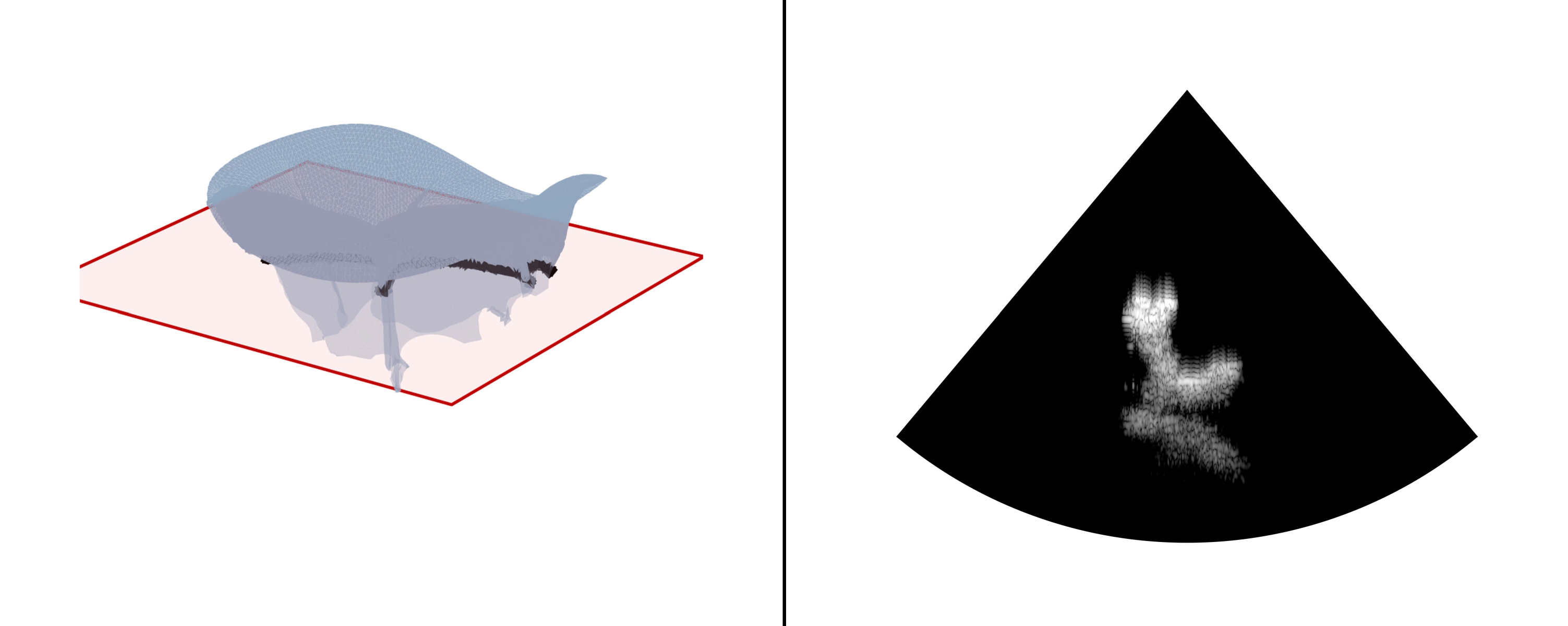}
        \caption{Top down view, $t = 1.0$ (fully closed).}
        \label{fig:top_t1}
    \end{subfigure}\hfill
    \begin{subfigure}[t]{0.48\linewidth}
        \centering
        \includegraphics[width=\linewidth]{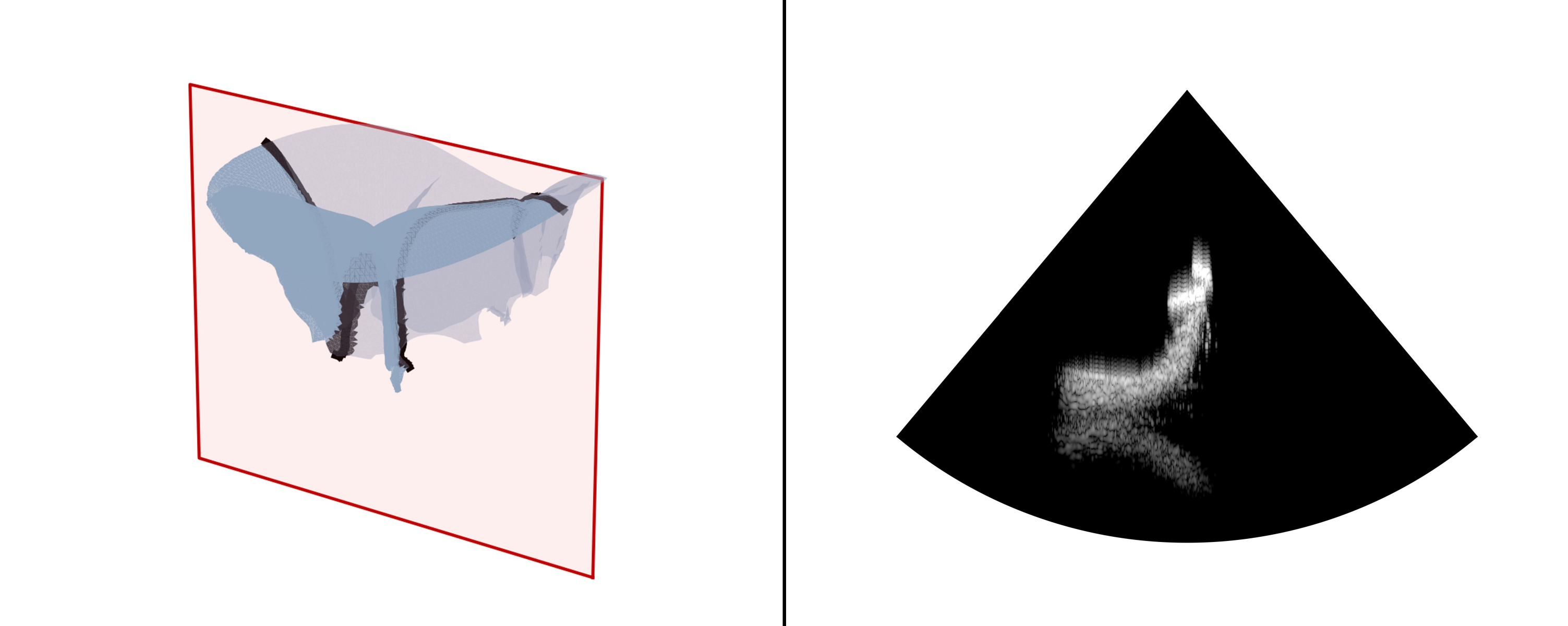}
        \caption{Side view, $t = 1.0$ (fully closed).}
        \label{fig:side_t1}
    \end{subfigure}

    \caption{Evolution of the right atrioventricular valve from the open to the closed state, shown from top down [\textbf{(a)}, \textbf{(c)}, \& \textbf{(e)}] and side views [\textbf{(b)}, \textbf{(d)}, \& \textbf{(f)}] at three representative time instances. The sequence illustrates the progressive reduction of the valve opening and the associated leaflet motion during closure, providing a more anatomically complex test case for the forward ultrasound simulation framework.}
    \label{fig:heart_valve}
\end{figure}

The second test case considers a right atrioventricular valve \cite{skwarek2006morphology} to demonstrate that the proposed simulation framework can capture more complex anatomical structures. Simulated B-mode scans are performed for both the open and closed valve states. The resulting images and corresponding valve models are shown in Figure~\ref{fig:heart_valve}. For additional details on the construction of the right atrioventricular valve model, we refer the reader to Mathur \cite{Mathur2022_TTV}.

\subsection{Optimization Problems}
\label{sec:optimization_problems}

The optimization problems considered in this work are based on the hollow cylinder scene shown in Figure~\ref{fig:cylinder_forward}. Specifically, we consider two settings: optimization with respect to a simulated reference image and optimization with respect to a real-world ultrasound image.

\subsubsection{Simulated to Simulated}
\label{sec:sim_reference_opt}

The first optimization setting considers a simulated-reference problem in which the target image is generated by the same forward model used in the optimization. This provides a controlled setting for assessing whether the proposed end-to-end differentiable pipeline can recover known scene parameters from image-space supervision alone.

In this experiment, we optimize three parameters of the hollow cylinder scene: the cylinder radius \(R\), acoustic impedance \(Z\), and surface roughness \(\rho\). These parameters were selected because they influence distinct aspects of the simulated B-mode image. The radius \(R\) primarily controls the geometric location and extent of the scattering interfaces, while \(Z\) and \(\rho\) affect echo strength and angular scattering behavior.

The target image is a simulated B-mode image generated using known ground-truth parameters. Starting from an initial guess, the optimization minimizes the weighted image-space \(\ell_1+\ell_2\) objective introduced in Section~\ref{sec:losses}, with gradients propagated through the full pipeline, including ray tracing, beamforming, and post-processing. Because the reference is synthetic and the underlying parameters are known, this experiment makes it possible to directly evaluate parameter recovery, loss convergence, and the consistency of the differentiable formulation.

\begin{figure}[H]
    \centering
    \begin{subfigure}[b]{0.33\linewidth}
        \centering
        \includegraphics[width=\linewidth]{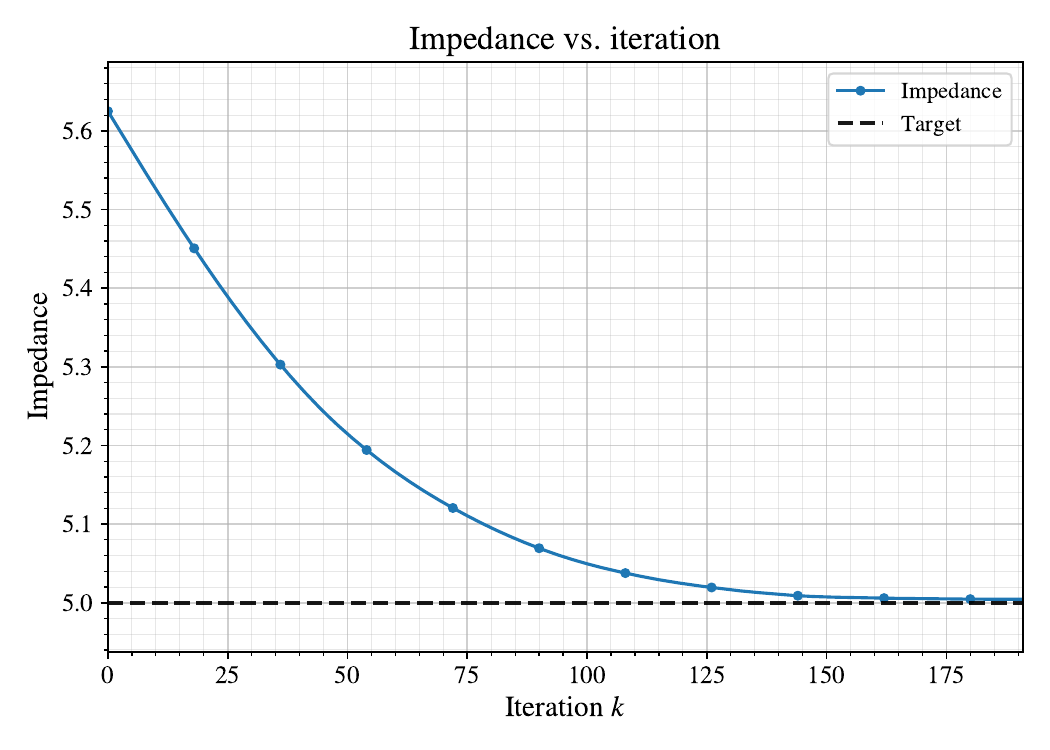}
        \caption{Evolution of impedance \(Z\).}
        \label{fig:sim_impedance}
    \end{subfigure}\hfill
    \begin{subfigure}[b]{0.33\linewidth}
        \centering
        \includegraphics[width=\linewidth]{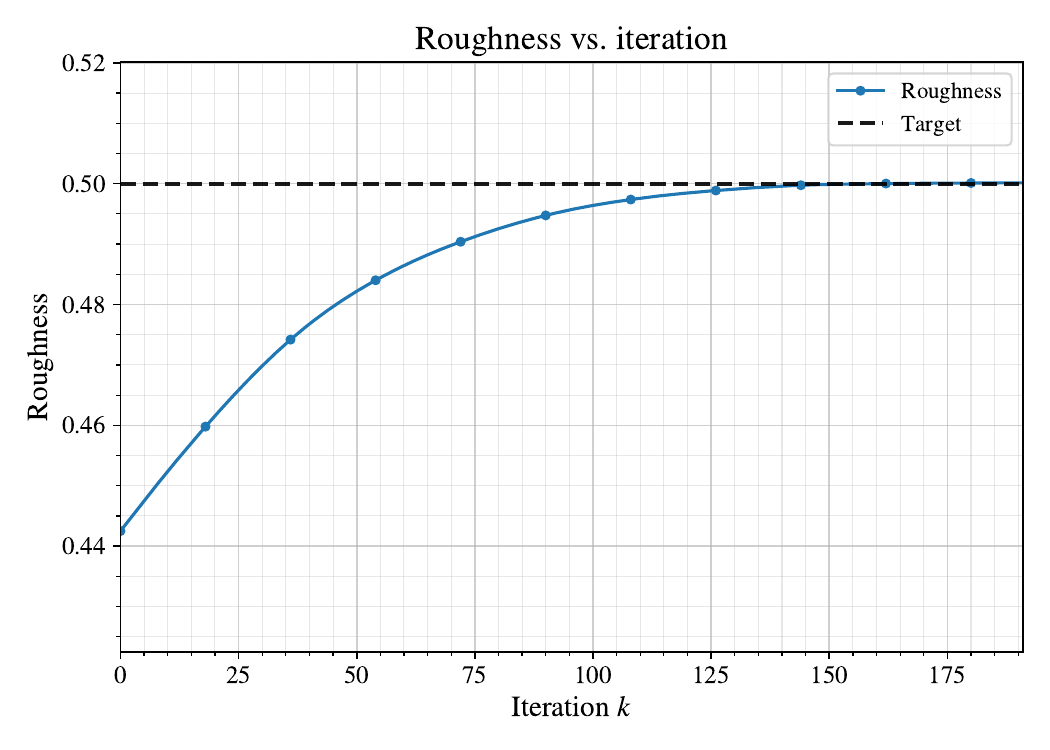}
        \caption{Evolution of roughness \(\rho\).}
        \label{fig:sim_roughness}
    \end{subfigure}\hfill
    \begin{subfigure}[b]{0.33\linewidth}
        \centering
        \includegraphics[width=\linewidth]{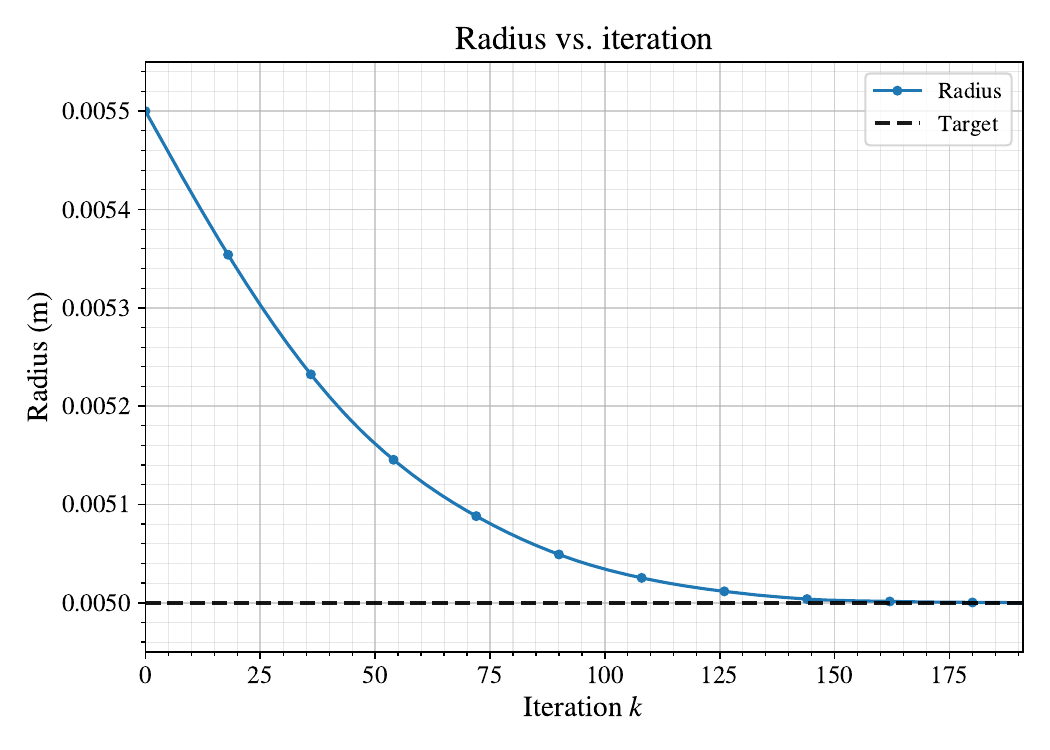}
        \caption{Evolution of radius \(R\).}
        \label{fig:sim_radius}
    \end{subfigure}
    \caption{Evolution of the optimized cylinder parameters in the simulated-reference experiment. The impedance \(Z\), roughness \(\rho\), and radius \(R\) progressively approach stable values consistent with the simulated target, demonstrating successful parameter recovery in the controlled inverse setting.}
    \label{fig:sim_param_histories}
\end{figure}

We report the evolution of the optimized parameters and the corresponding loss history (see Figure~\ref{fig:sim_param_histories}). We also compare automatic-differentiation gradients with finite-difference estimates to assess the consistency of the computed sensitivities with the underlying forward model; this comparison is discussed in more detail later in this section. In addition, we show the initial simulation, the optimized final simulation, and the reference image for qualitative comparison.

\begin{figure}[H]
    \centering
    \includegraphics[width=0.6\linewidth]{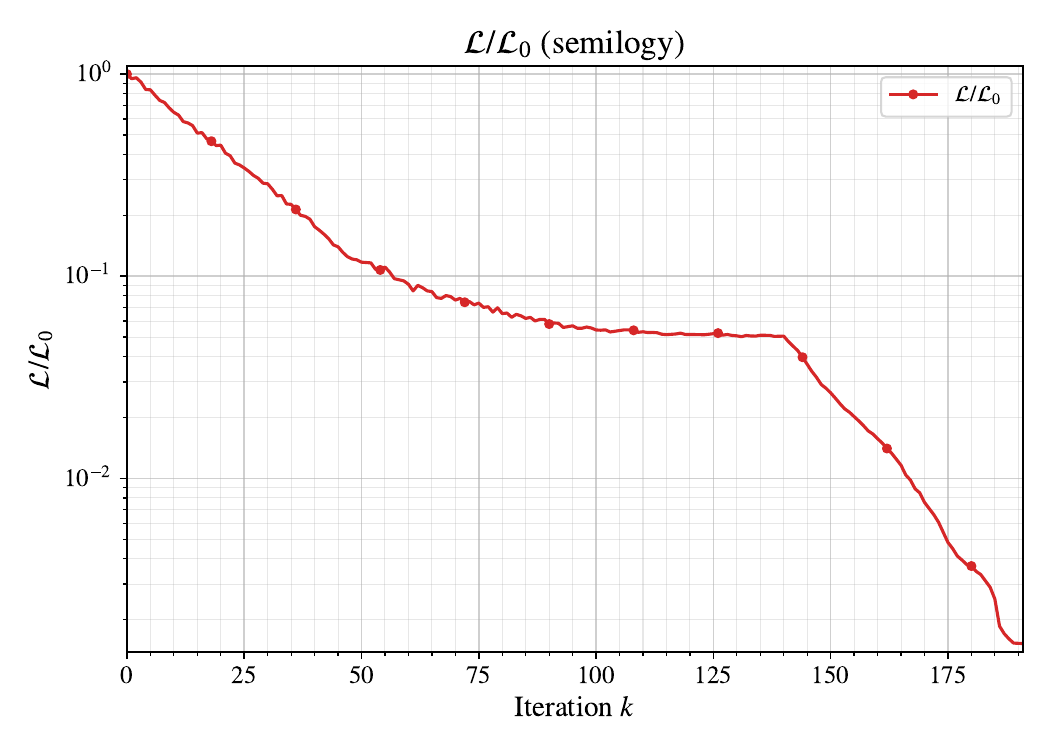}
    \caption{History of the weighted \(\ell_1+\ell_2\) image-space loss for the simulated-reference optimization problem. The loss decreases rapidly over the course of the optimization and approaches a near-plateau at a very low value, indicating that the differentiable pipeline successfully identifies a parameter set that closely reproduces the simulated reference image.}
    \label{fig:sim_loss}
\end{figure}

\begin{figure}[hbtp]
    \centering

    \begin{subfigure}[b]{0.33\linewidth}
        \centering
        \includegraphics[width=\linewidth]{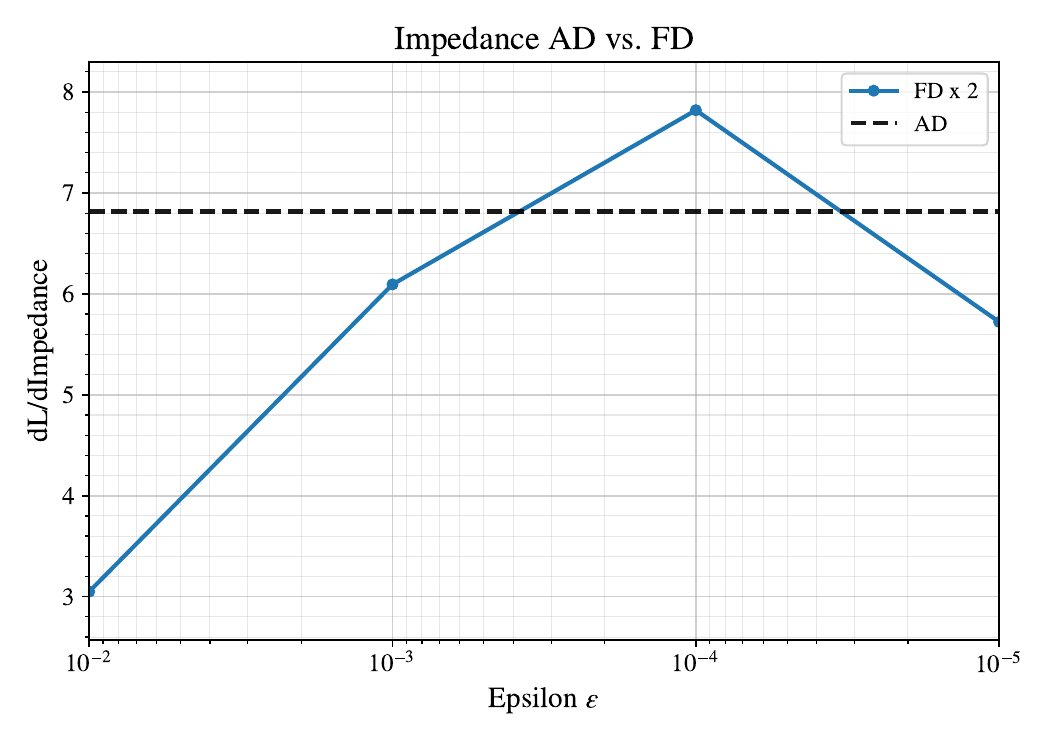}
        \caption{$Z$: FD and AD vs. $\epsilon$.}
        \label{fig:grad_raw_Z}
    \end{subfigure}\hfill
    \begin{subfigure}[b]{0.33\linewidth}
        \centering
        \includegraphics[width=\linewidth]{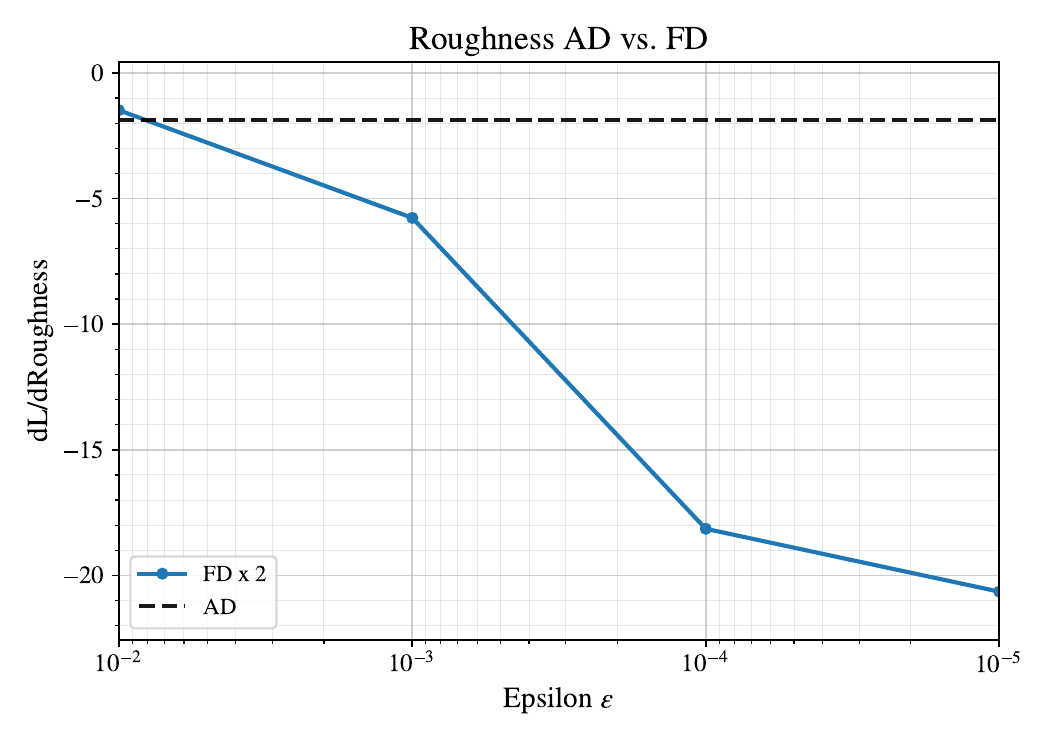}
        \caption{$\rho$: FD and AD vs. $\epsilon$.}
        \label{fig:grad_raw_rho}
    \end{subfigure}\hfill
    \begin{subfigure}[b]{0.33\linewidth}
        \centering
        \includegraphics[width=\linewidth]{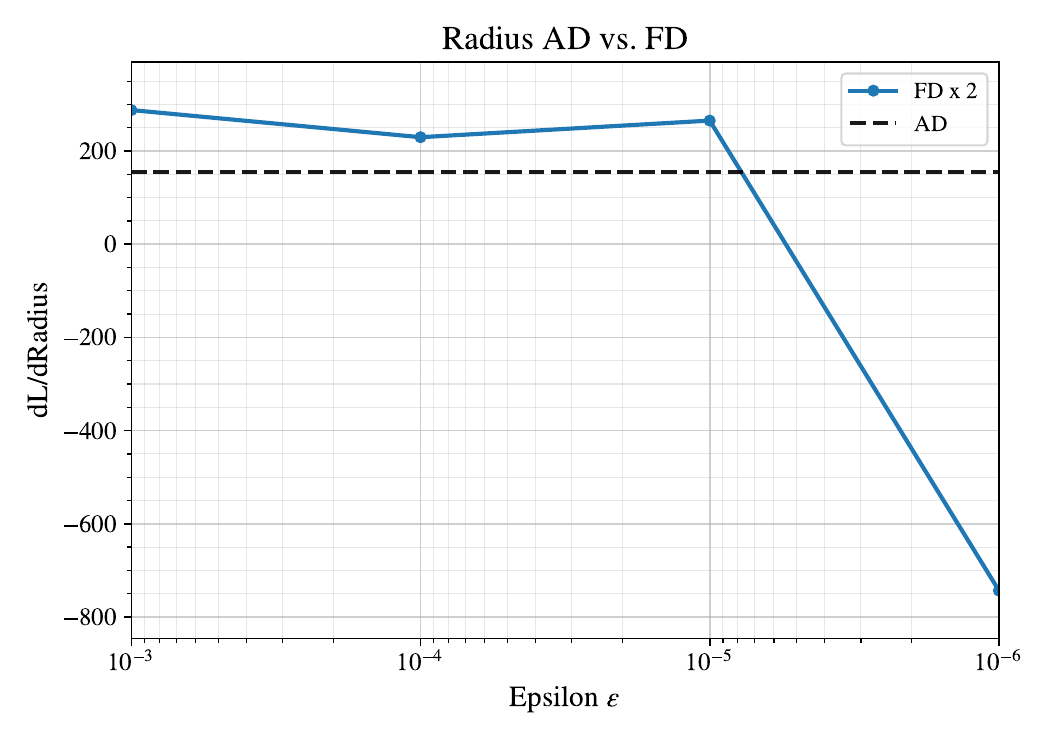}
        \caption{$R$: FD and AD vs. $\epsilon$.}
        \label{fig:grad_raw_R}
    \end{subfigure}

    \vspace{0.8em}

    \begin{subfigure}[b]{0.33\linewidth}
        \centering
        \includegraphics[width=\linewidth]{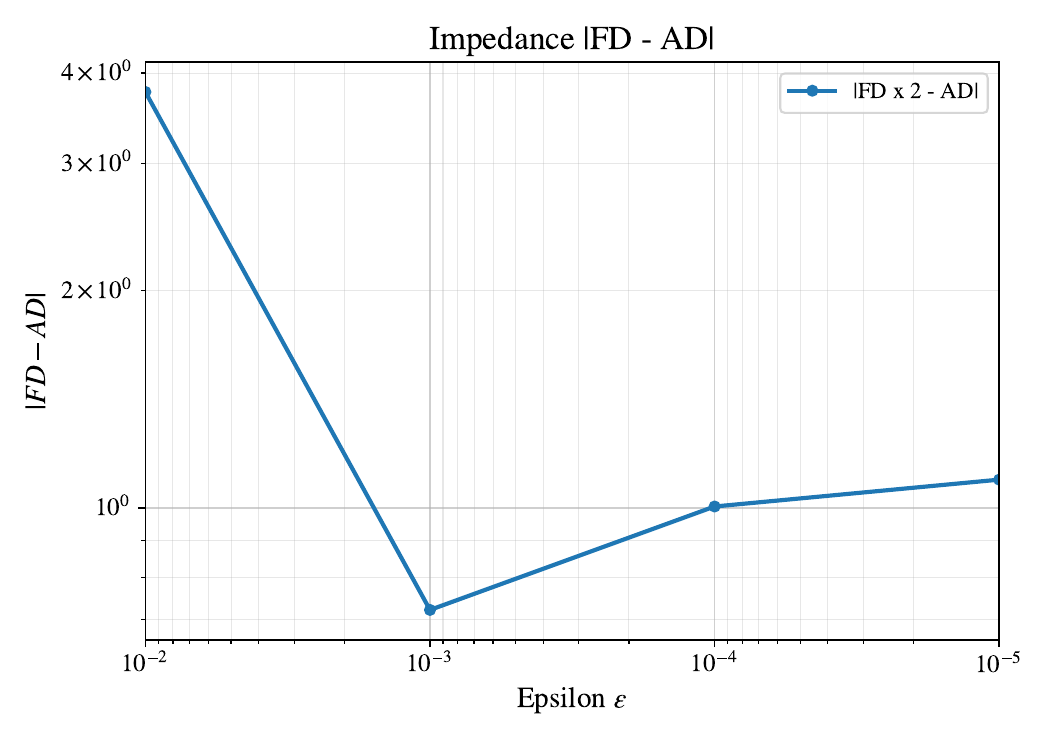}
        \caption{$Z$: $\lVert g_{\mathrm{FD}}-g_{\mathrm{AD}}\rVert$ vs. $\epsilon$.}
        \label{fig:grad_norm_Z}
    \end{subfigure}\hfill
    \begin{subfigure}[b]{0.33\linewidth}
        \centering
        \includegraphics[width=\linewidth]{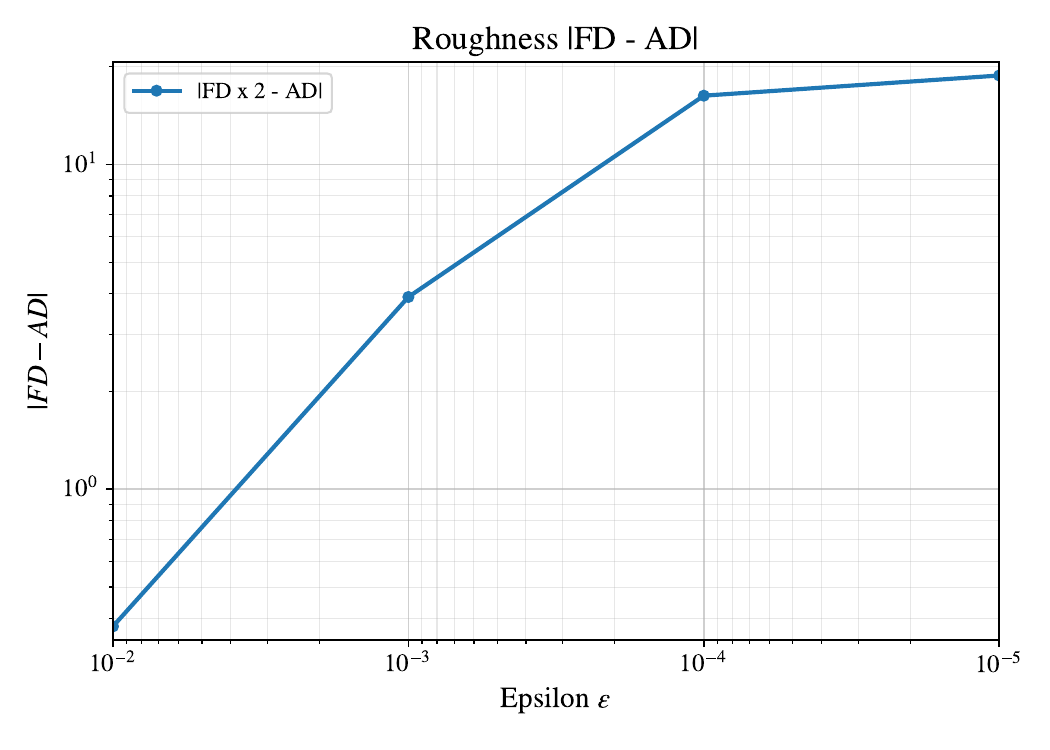}
        \caption{$\rho$: $\lVert g_{\mathrm{FD}}-g_{\mathrm{AD}}\rVert$ vs. $\epsilon$.}
        \label{fig:grad_norm_rho}
    \end{subfigure}\hfill
    \begin{subfigure}[b]{0.33\linewidth}
        \centering
        \includegraphics[width=\linewidth]{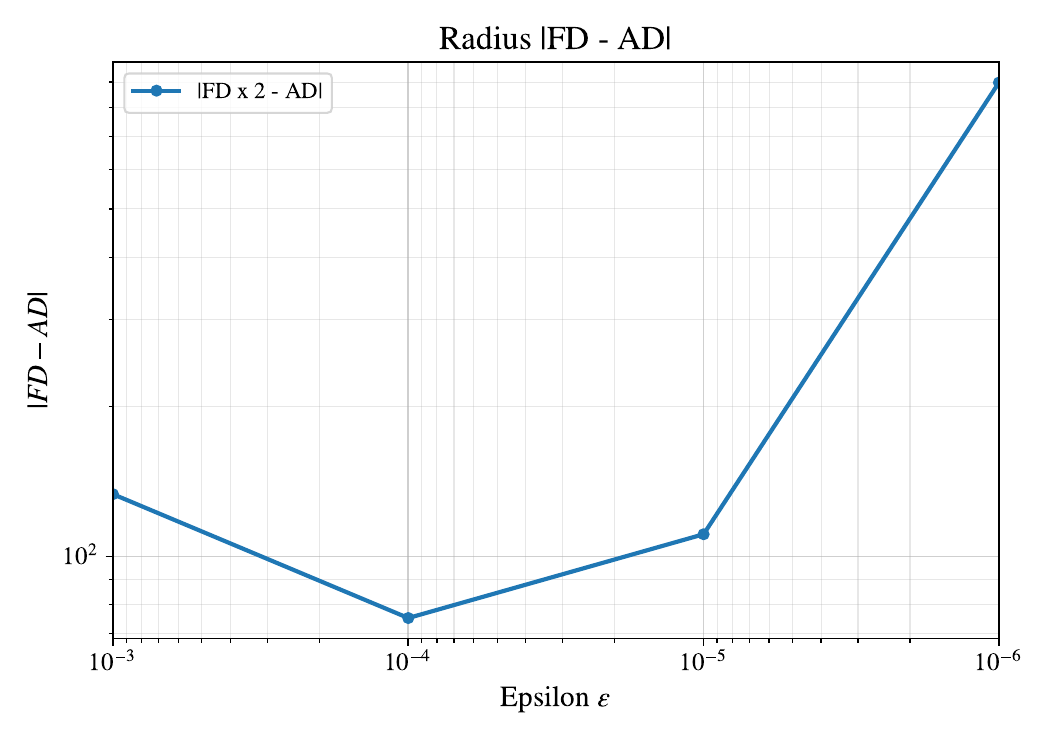}
        \caption{$R$: $\lVert g_{\mathrm{FD}}-g_{\mathrm{AD}}\rVert$ vs. $\epsilon$.}
        \label{fig:grad_norm_R}
    \end{subfigure}

    \caption{Automatic-differentiation (AD) and finite-difference (FD) gradient comparisons for the three optimized parameters in the simulated-reference cylinder problem. The top row reports the AD and FD gradient values for impedance \(Z\), roughness \(\rho\), and radius \(R\) as functions of the FD step size \(\epsilon\), and the bottom row reports the corresponding discrepancy \(\lVert g_{\mathrm{FD}}-g_{\mathrm{AD}}\rVert\). The close agreement observed over an intermediate range of \(\epsilon\) supports the consistency of the AD sensitivities with the underlying forward model, while the divergence at very small and very large perturbation sizes reflects the expected limitations of FD validation due to numerical cancellation and truncation error.}
    \label{fig:sim_grad_compare}
\end{figure}

From Figures~\ref{fig:sim_loss} and~\ref{fig:sim_grad_compare}, the radius exhibits the strongest influence on the loss, with gradient magnitudes that are orders of magnitude larger than those of the other parameters. The next most influential parameter is impedance, followed by roughness. This ordering persists in later examples and is also physically intuitive.

The cylinder radius directly controls the target geometry, specifically the available reflecting surface area and the locations at which reflections and transmissions occur, so it naturally has the strongest effect on both the \(\ell_1\) and \(\ell_2\) components of the loss. Impedance primarily modulates how energy is partitioned between reflected and transmitted components, strongly affecting echo amplitudes and the behavior of secondary paths. Finally, roughness governs angular scattering, so its effect on the loss is typically more indirect and therefore weaker by comparison.

The gradient comparisons in Figure~\ref{fig:sim_grad_compare} provide an additional verification of the differentiable formulation. For each parameter, we sweep the finite-difference step size \(\epsilon\) and compare the resulting finite-difference estimate with the gradient obtained by automatic differentiation. In each case, the two approaches show close agreement for a suitable range of \(\epsilon\), indicating that the automatic-differentiation gradients are consistent with finite-difference calculations.

\begin{figure}[htbp]
    \centering
    \begin{subfigure}[b]{0.32\linewidth}
        \centering
        \includegraphics[width=\linewidth]{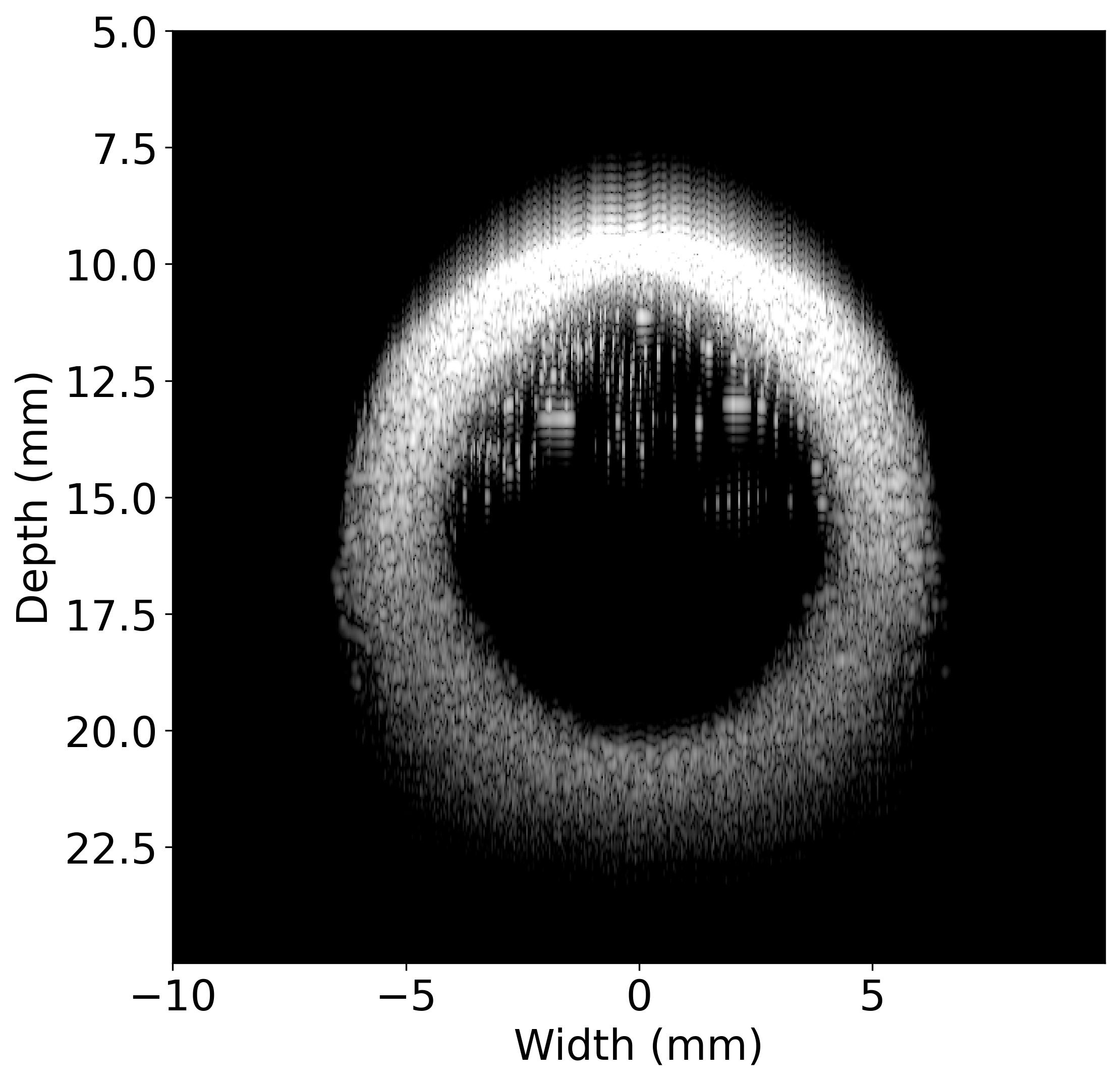}
        \caption{Initial simulated image.}
        \label{fig:sim_initial_image}
    \end{subfigure}\hfill
    \begin{subfigure}[b]{0.32\linewidth}
        \centering
        \includegraphics[width=\linewidth]{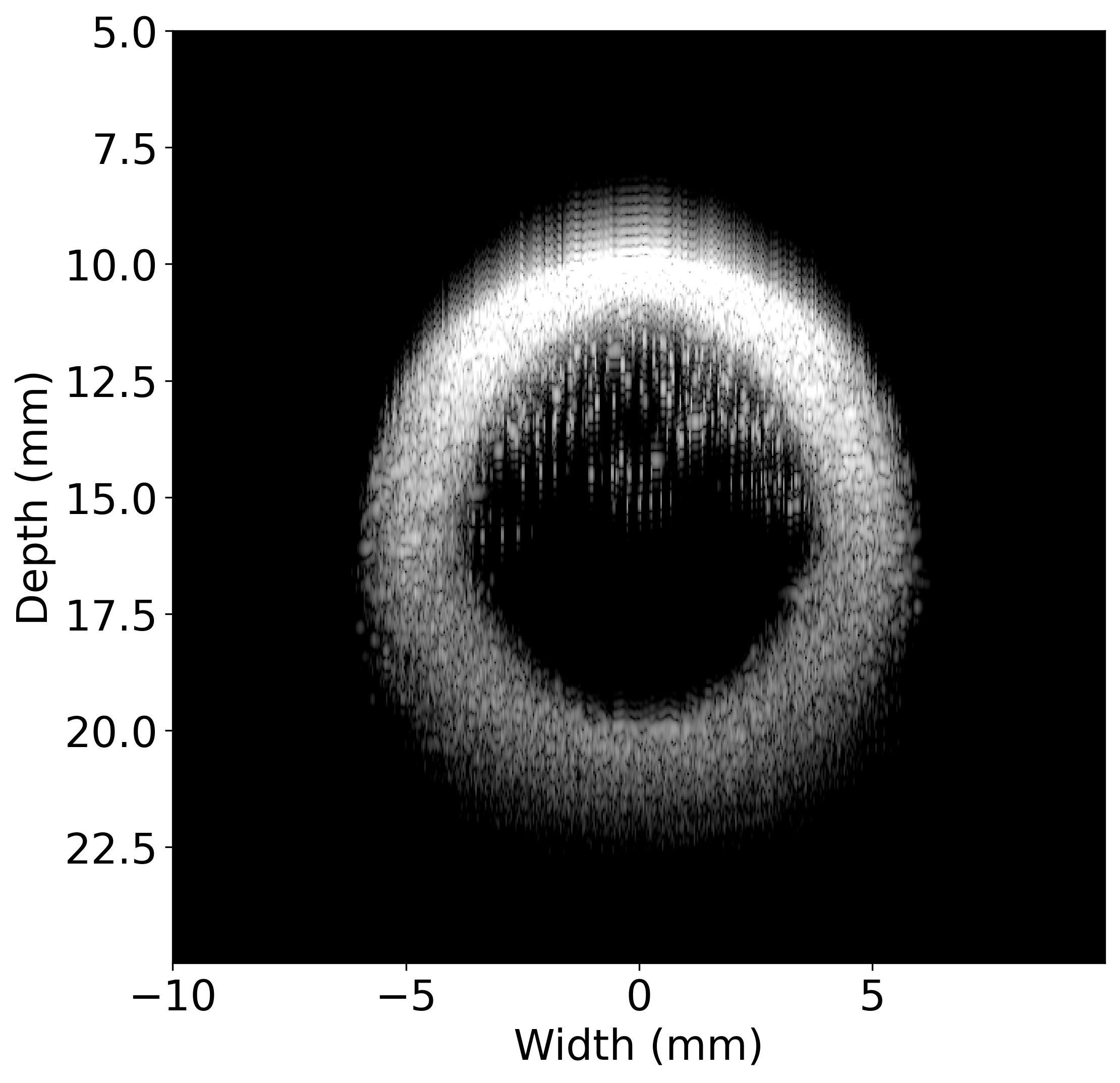}
        \caption{Optimized simulated image.}
        \label{fig:sim_final_image}
    \end{subfigure}\hfill
    \begin{subfigure}[b]{0.32\linewidth}
        \centering
        \includegraphics[width=\linewidth]{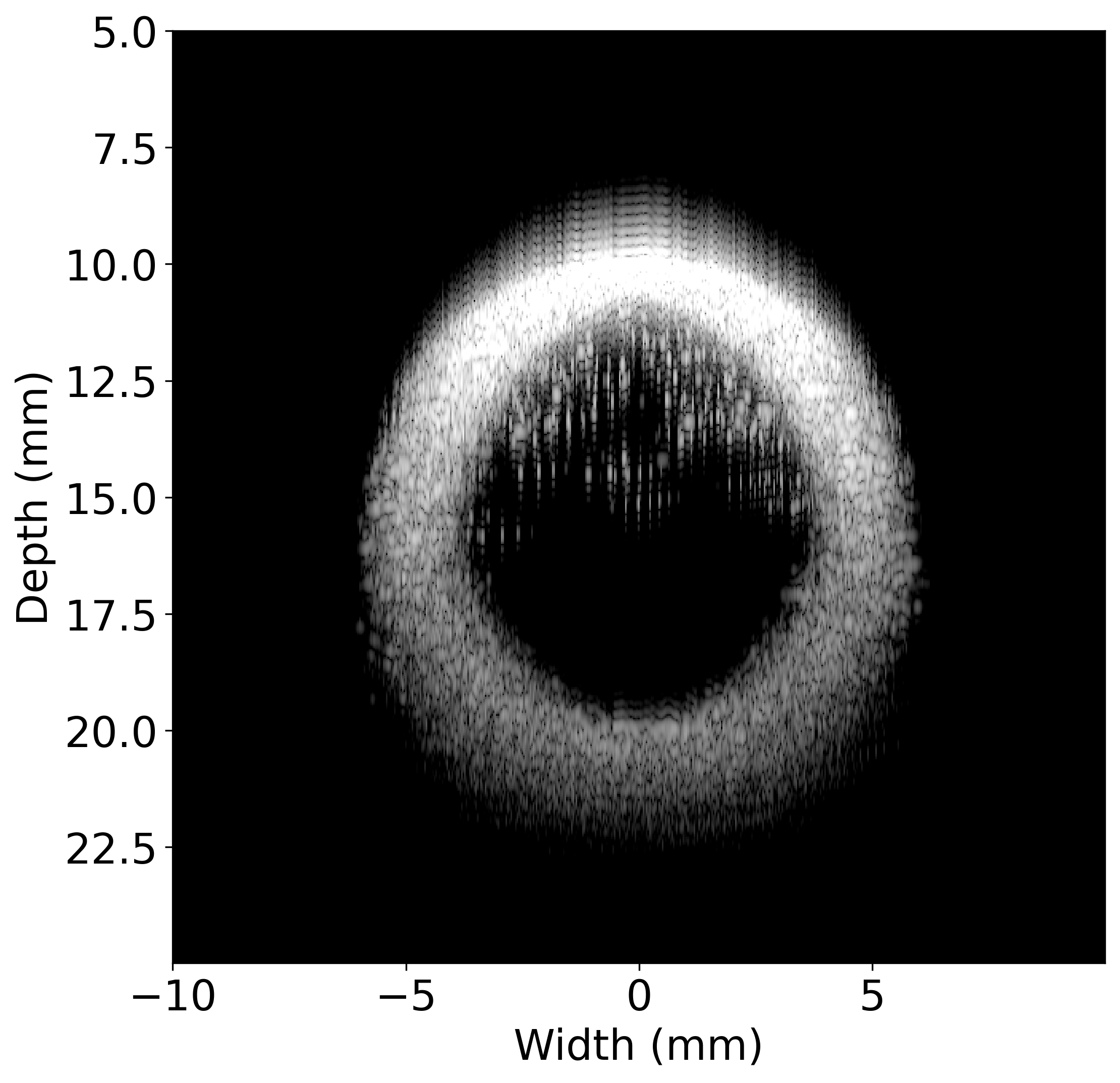}
        \caption{Reference image.}
        \label{fig:sim_reference_image}
    \end{subfigure}
    \caption{Comparison of the initial simulated image, the optimized simulated image, and the simulated reference image for the simulated-reference optimization problem. Relative to the initial configuration, the optimized result achieves near-complete agreement with the reference image, indicating that the differentiable pipeline successfully recovers a parameter set that reproduces the dominant geometric and intensity features of the target.}
    \label{fig:sim_image_compare}
\end{figure}

Figure~\ref{fig:sim_image_compare} shows the initial simulated image, the optimized final image, and the reference image. The optimized image is visually nearly indistinguishable from the reference image, further supporting the success of the optimization. This qualitative agreement is also consistent with the observed loss reduction and the recovery of parameters close to their ground-truth values.

\subsubsection{Simulated to Real-World}
\label{sec:reals_bmode_opt}

\begin{figure}[htpb]
    \centering
    \begin{subfigure}[b]{0.32\linewidth}
        \centering
        \includegraphics[width=\linewidth]{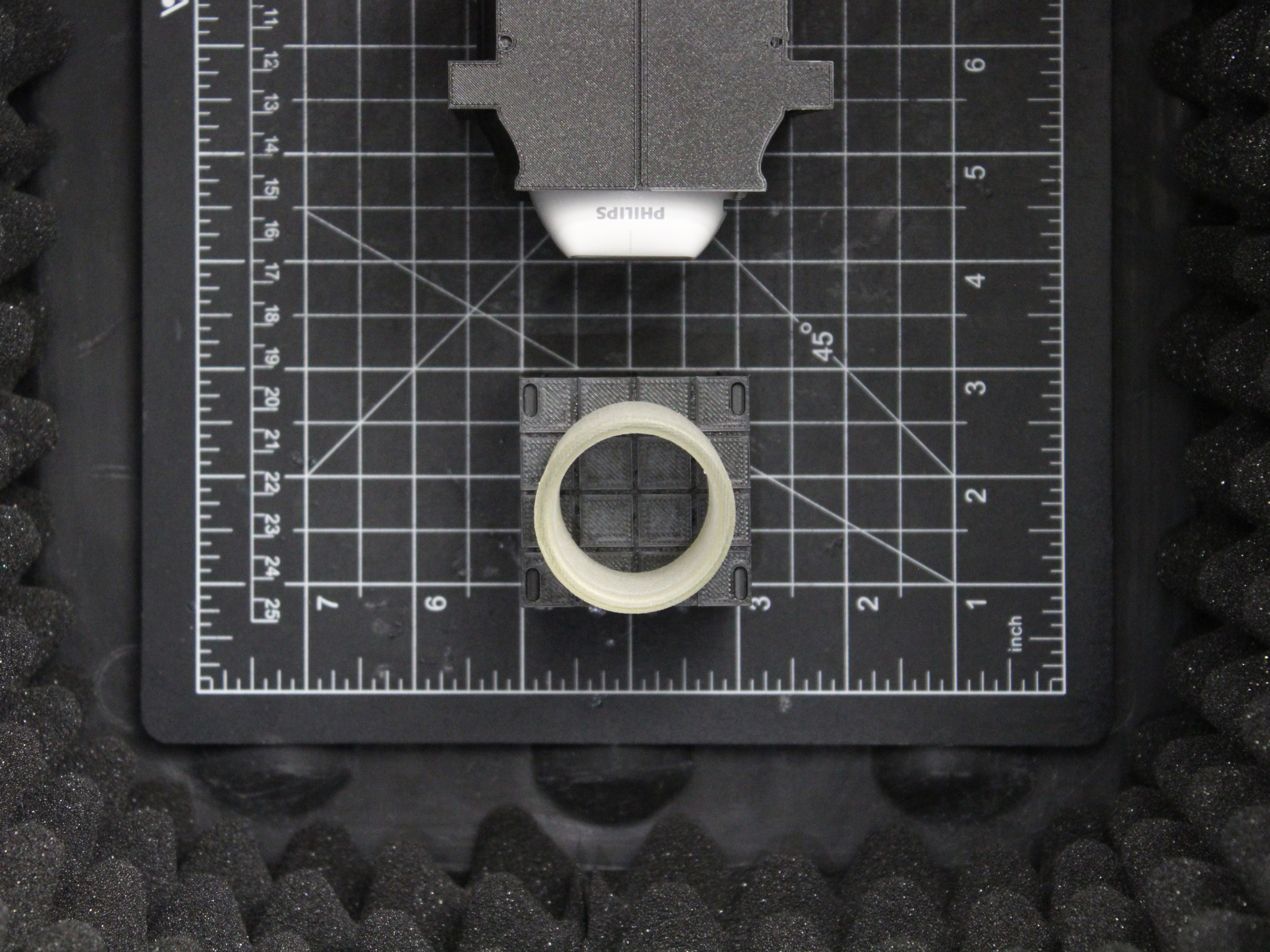}
        \caption{Experimental Setup.}
        \label{fig:experimental_setup}
    \end{subfigure}
    \begin{subfigure}[b]{0.32\linewidth}
        \centering
        \includegraphics[width=\linewidth]{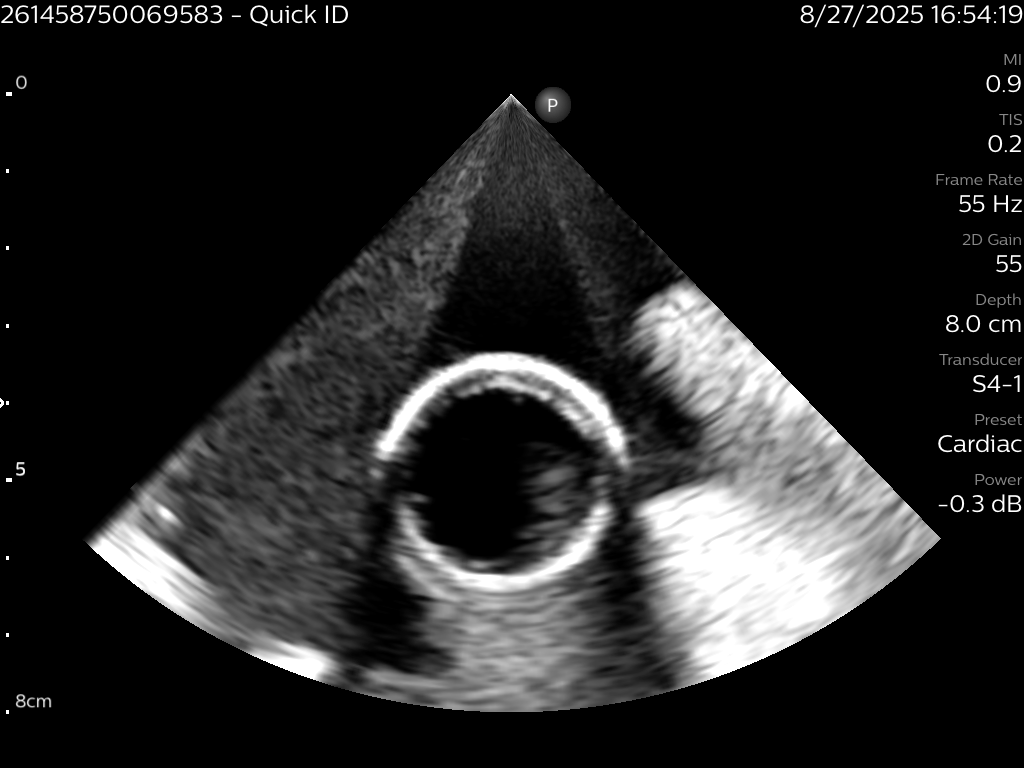}
        \caption{Raw experimental image.}
        \label{fig:real_world_example}
    \end{subfigure}
    \begin{subfigure}[b]{0.32\linewidth}
        \centering
        \includegraphics[width=\linewidth]{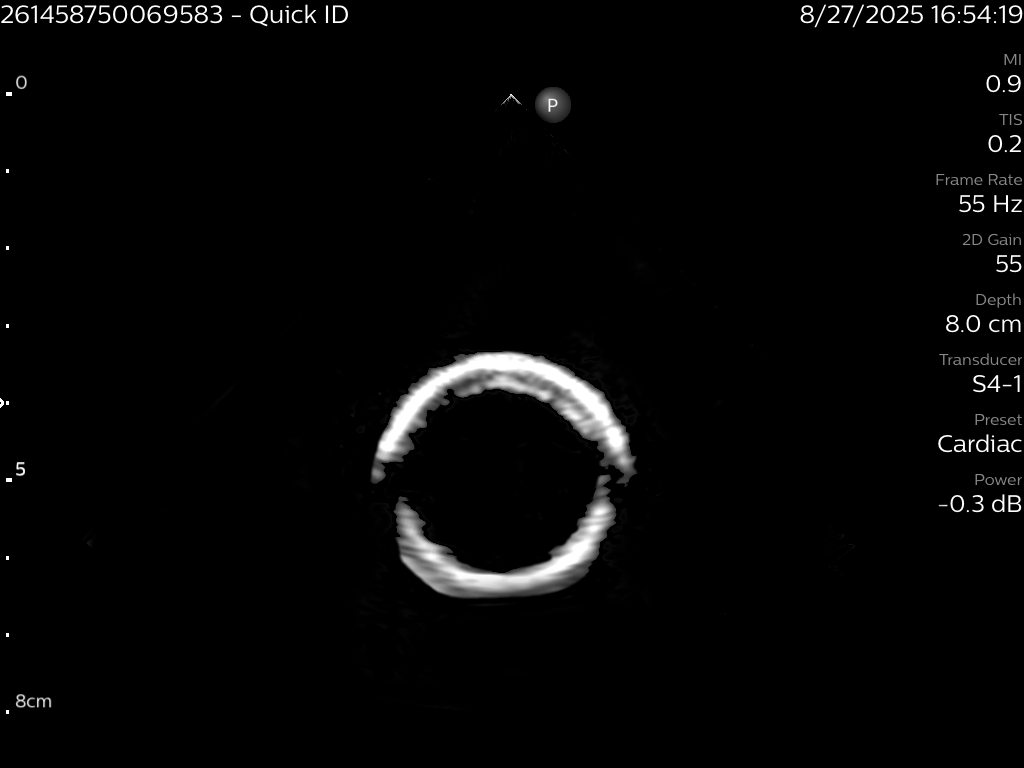}
        \caption{Post-processed experimental image.}
        \label{fig:real_world_cleaned_up}
    \end{subfigure}
    \caption{(a) Experimental setup used to capture B-mode images of a 3D-printed cylinder. The cylinder and Philips Lumify S4-1 probe are submerged in an acoustic damping foam-lined tub of water, and raised 2.25" from the bottom to minimize acoustic reflections. Note, the ultrasound imaging plane is parallel to the image and water is removed for clarity. (b) Unprocessed experimental B-mode image of the $R_{\mathrm{exp}} = 20$ mm by \(t_{\mathrm{exp}} = 3\) mm cylinder placed 4 cm from the probe. (c) The post-processed image used as the real-world target objective in the simulation-to-real optimization. We cropped the image to include only the main acoustic scattering effects of the cylinder and ignore noise from the surrounding fluid.}
    \label{fig:real_world_pair}
\end{figure}

The second optimization setting considers a simulation-to-real inverse problem in which a simulated B-mode image is matched to an experimentally acquired ultrasound image. Figure~\ref{fig:real_world_pair} shows the experimental setup, the original target image, and the post-processed image used in the optimization. This post-processing is introduced to suppress noise and emphasize the dominant scattering features that can be represented more reliably by the simulation model.

The goal of this experiment is to assess whether the proposed pipeline can identify an effective set of simulation parameters that reproduces the salient characteristics of an observed real measurement. As in the simulated-reference case, we optimize three parameters of the hollow-cylinder scene: the cylinder radius \(R\), acoustic impedance \(Z\), and surface roughness \(\rho\). These parameters were selected because they influence complementary aspects of the simulated image: \(R\) primarily controls the interface geometry, while \(Z\) and \(\rho\) mainly affect echo amplitude and angular scattering behavior.

The setup used to acquire the B-mode image of a cylinder is shown in the Figure \ref{fig:real_world_pair}(a). We placed a 12" x 9" gridded cutting mat in a 20" x 13" x 5" polypropylene tub, lined the tub with acoustic dampening foam (1.5" egg crate foam, WVOVW), and filled the tub with water. We 3D-printed a 40 mm diameter by 50 mm tall by 3 mm thick cylinder of "highly contractile myocardium" mimicking material on a Stratasys J750 Digital Anatomy 3D-printer (Stratasys, Eden Prairie, MN, USA, see \cite{bechtel2025well} for details) in order to approximate the material properties of tissue. Next, we submerged and centered the cylinder on a 2 .25" 3D-printed stand in the water-filled tub, and similarly centered the ultrasound probe on a 2.25" 3D-printed stand placed 4 cm from the center of the cylinder. The cylinder and probe were raised to minimize acoustic reflections from the bottom of the tub. Lastly, 2D B-mode ultrasound images were captured through the cross-section of the cylinder using a Philips Lumify S4-1 probe (Philips Healthcare, Amsterdam, Netherlands), as shown in Figure \ref{fig:real_world_pair}(b). Unlike the simulated-reference case, the experimental target does not provide a complete ground-truth parameter set. The primary known geometric quantity is the physical ring radius, \(R_{\mathrm{exp}} = 20\) mm and finite ring thickness of \(t_{\mathrm{exp}} = 3\) mm. However, the surface roughness and and acoustic impedance of the cylinder are unknown.

For image comparison, the experimental image was rescaled by a factor of \(3.5/20 = 0.175\), yielding an effective radius of \(R_{\mathrm{eff}} = 3.5\) mm. This rescaling was performed to obtain improved cropping, image arrangement, and geometric agreement with the simulated image domain, thereby enabling a more consistent visual correspondence between the dominant experimental and simulated features. The rescaled image is therefore treated as an effective reference for the optimization. This comparison is further supported by the approximate scale-invariant behavior of the model when the scene geometry and acquisition settings are scaled consistently.

At the same time, the experimental ring is not an ideal infinitesimally thin interface. Its nonzero thickness introduces additional scattering structure and broadens the observed boundary response, which can contribute to residual mismatch relative to the simplified simulated model. Moreover, because the experimental target does not uniquely determine all scene parameters, the recovered values of \(Z\) and \(\rho\) should be interpreted as effective parameters that reproduce the dominant observed image features rather than as uniquely identified physical material properties. Accordingly, evaluation in this setting focuses on reduction of the weighted image-space \(\ell_1+\ell_2\) objective together with qualitative agreement between the optimized simulation and the processed experimental target. We again report the evolution of the optimized parameters, the loss history, and the associated gradients.

\begin{figure}[htbp]
    \centering
    \begin{subfigure}[b]{0.33\linewidth}
        \centering
        \includegraphics[width=\linewidth]{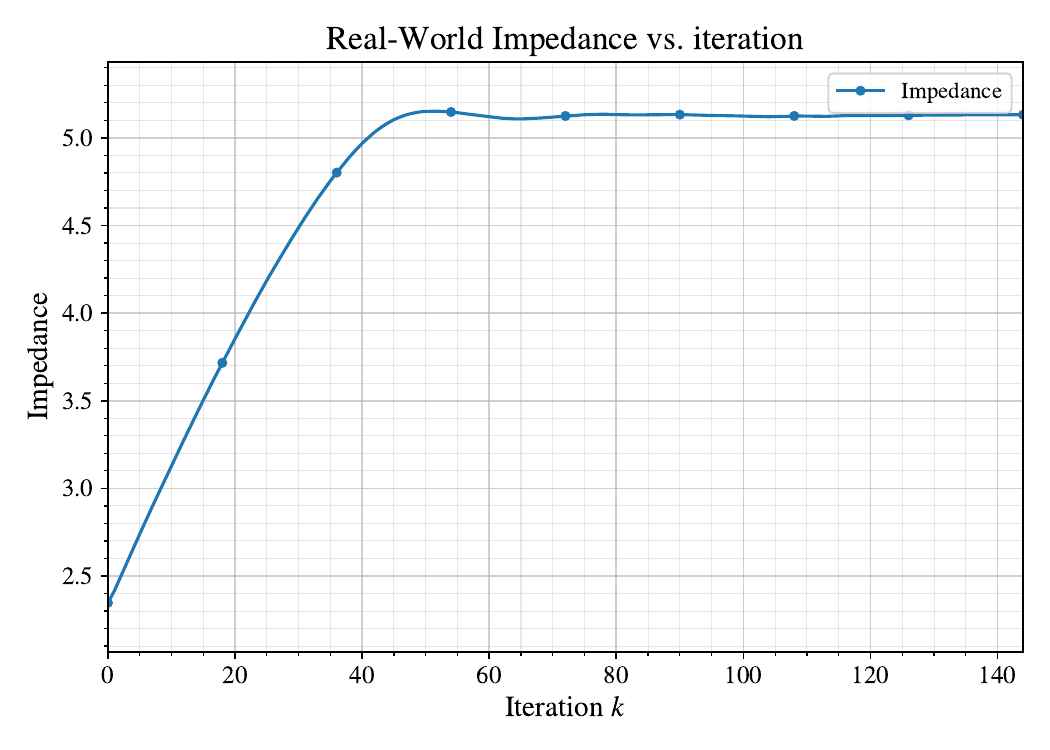}
        \caption{Evolution of impedance \(Z\).}
        \label{fig:real_impedance}
    \end{subfigure}\hfill
    \begin{subfigure}[b]{0.33\linewidth}
        \centering
        \includegraphics[width=\linewidth]{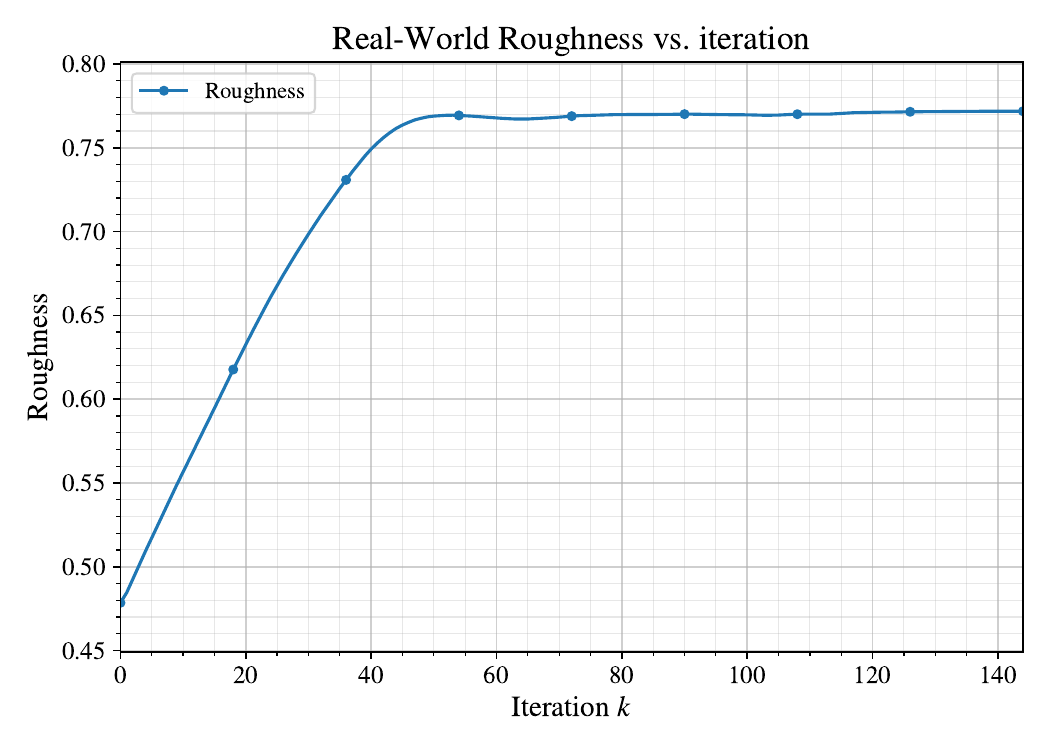}
        \caption{Evolution of roughness \(\rho\).}
        \label{fig:real_roughness}
    \end{subfigure}\hfill
    \begin{subfigure}[b]{0.33\linewidth}
        \centering
        \includegraphics[width=\linewidth]{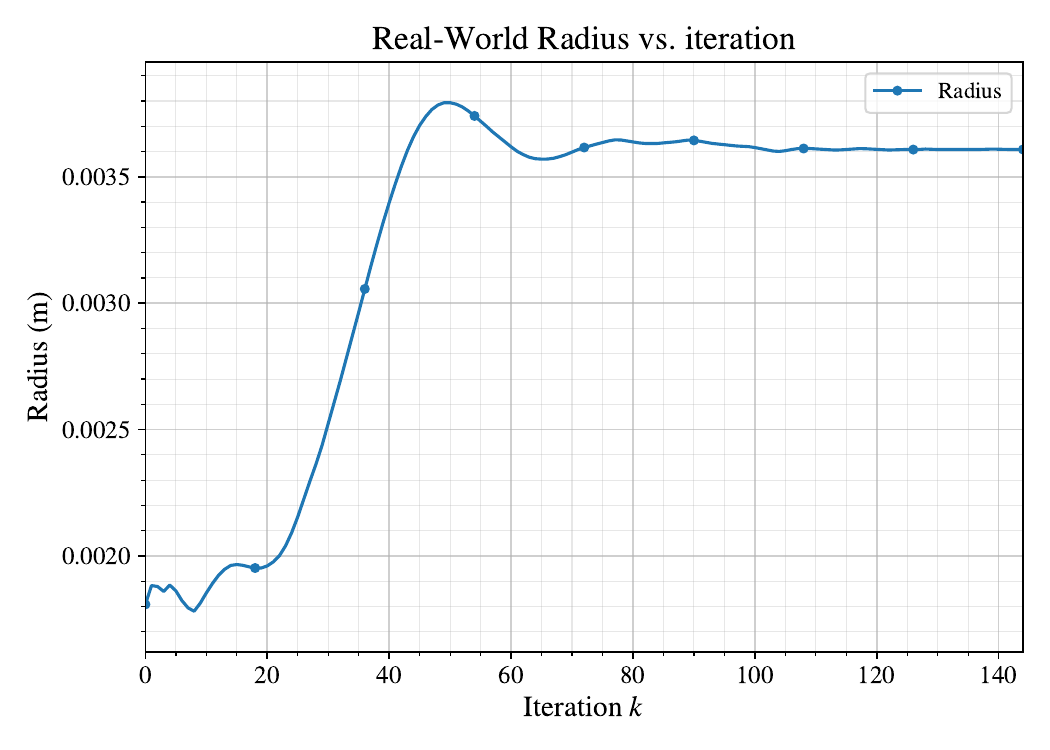}
        \caption{Evolution of radius \(R\).}
        \label{fig:real_radius}
    \end{subfigure}
    \caption{Evolution of the optimized cylinder parameters in the simulation-to-real experiment. The impedance \(Z\), roughness \(\rho\), and radius \(R\) all move away from their initial values and progressively stabilize, indicating convergence toward an effective parameter set that improves agreement between the simulated and experimental B-mode images.}
    \label{fig:real_param_histories}
\end{figure}

\begin{figure}[hbtp]
    \centering
    \includegraphics[width=0.6\linewidth]{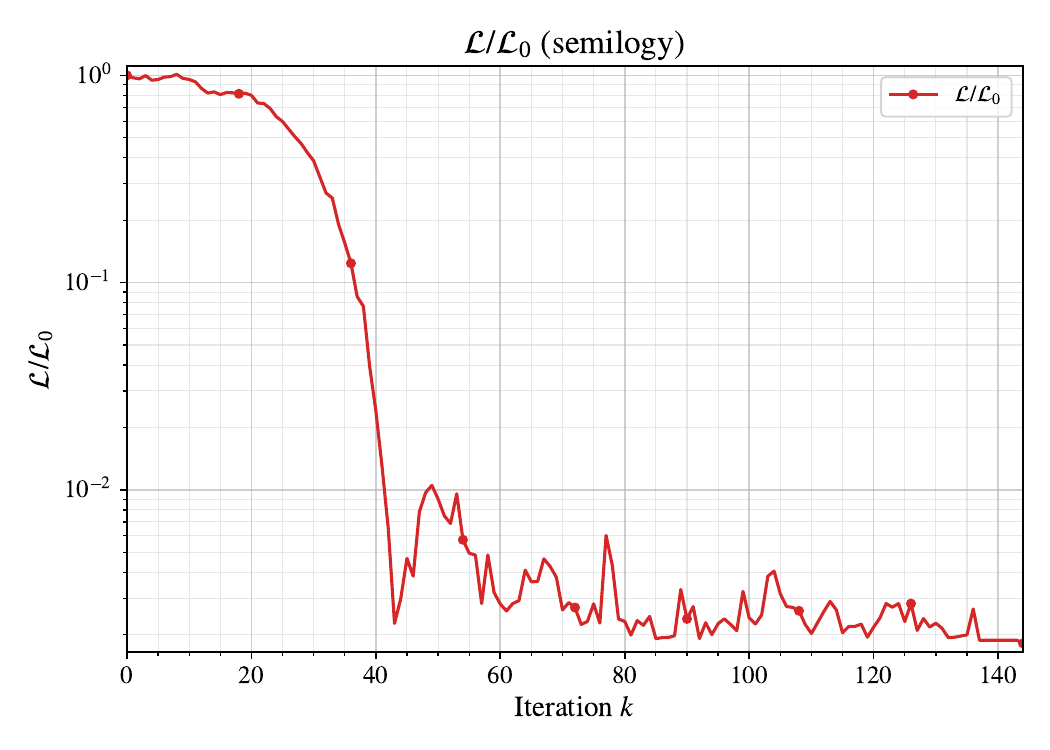}
    \caption{History of the weighted \(\ell_1+\ell_2\) image-space loss for the simulation-to-real optimization problem. The loss exhibits a pronounced decrease during the early iterations, followed by gradual stabilization, indicating that the optimizer identifies a parameter regime with substantially improved agreement between the simulated and experimental B-mode images.}
    \label{fig:real_loss}
\end{figure}

\begin{figure}[hbtp]
    \centering

    \begin{subfigure}[b]{0.33\linewidth}
        \centering
        \includegraphics[width=\linewidth]{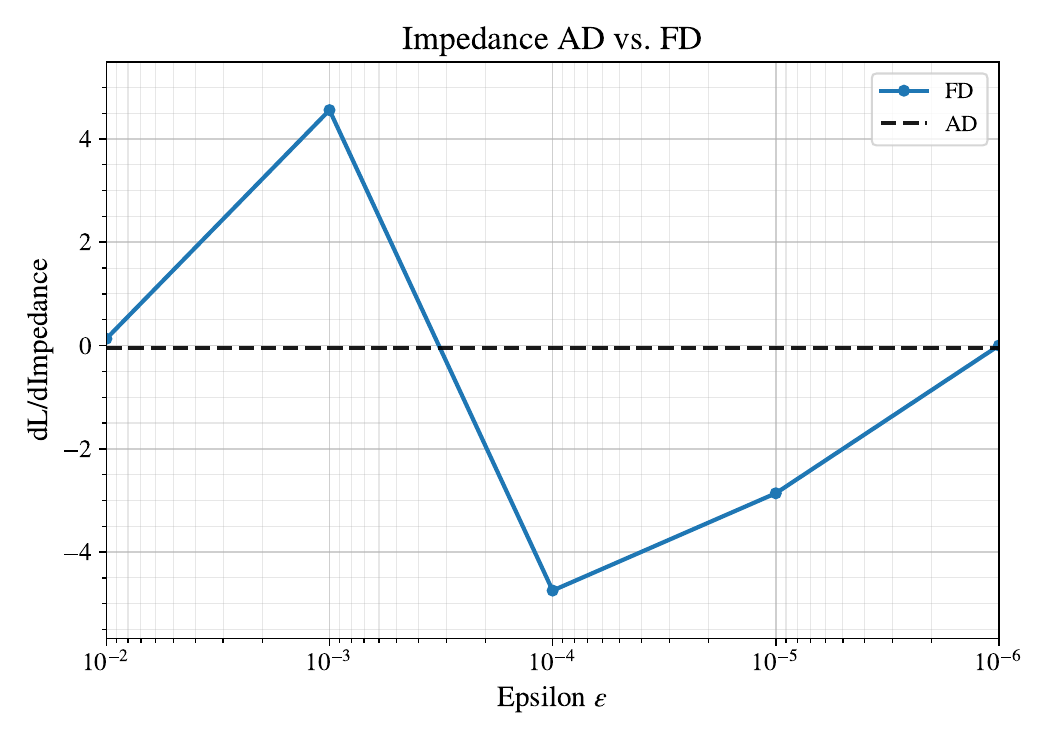}
        \caption{$Z$: FD and AD vs. $\epsilon$.}
        \label{fig:grad_raw_Z}
    \end{subfigure}\hfill
    \begin{subfigure}[b]{0.33\linewidth}
        \centering
        \includegraphics[width=\linewidth]{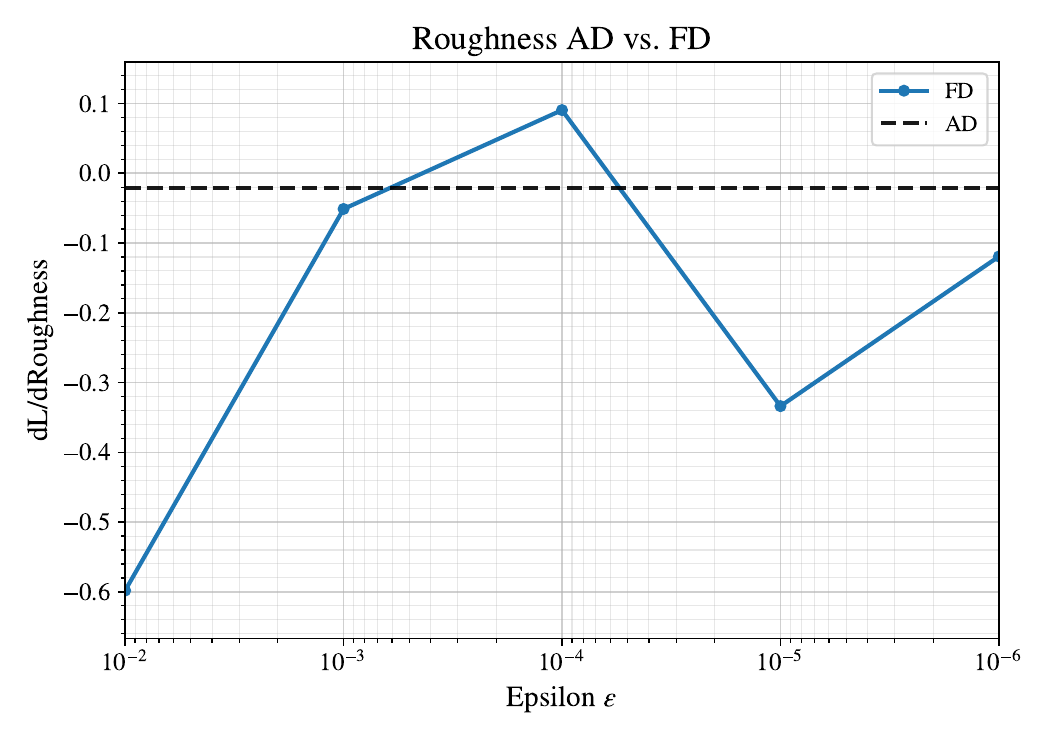}
        \caption{$\rho$: FD and AD vs. $\epsilon$.}
        \label{fig:grad_raw_rho}
    \end{subfigure}\hfill
    \begin{subfigure}[b]{0.33\linewidth}
        \centering
        \includegraphics[width=\linewidth]{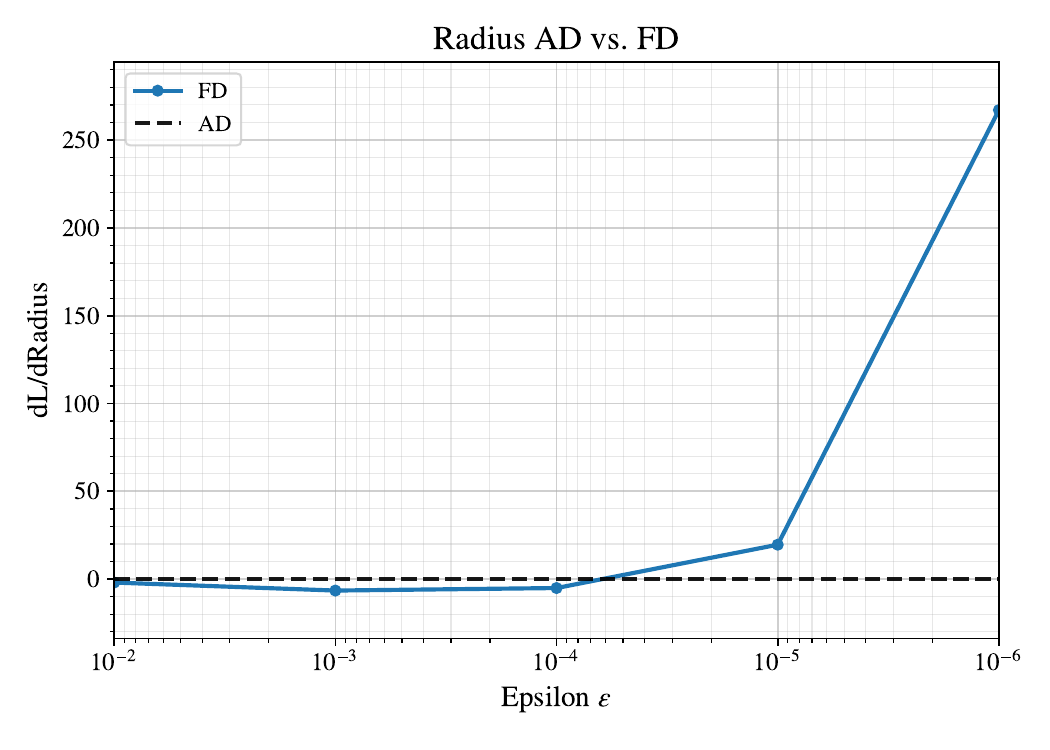}
        \caption{$R$: FD and AD vs. $\epsilon$.}
        \label{fig:grad_raw_R}
    \end{subfigure}

    \vspace{0.8em}

    \begin{subfigure}[b]{0.33\linewidth}
        \centering
        \includegraphics[width=\linewidth]{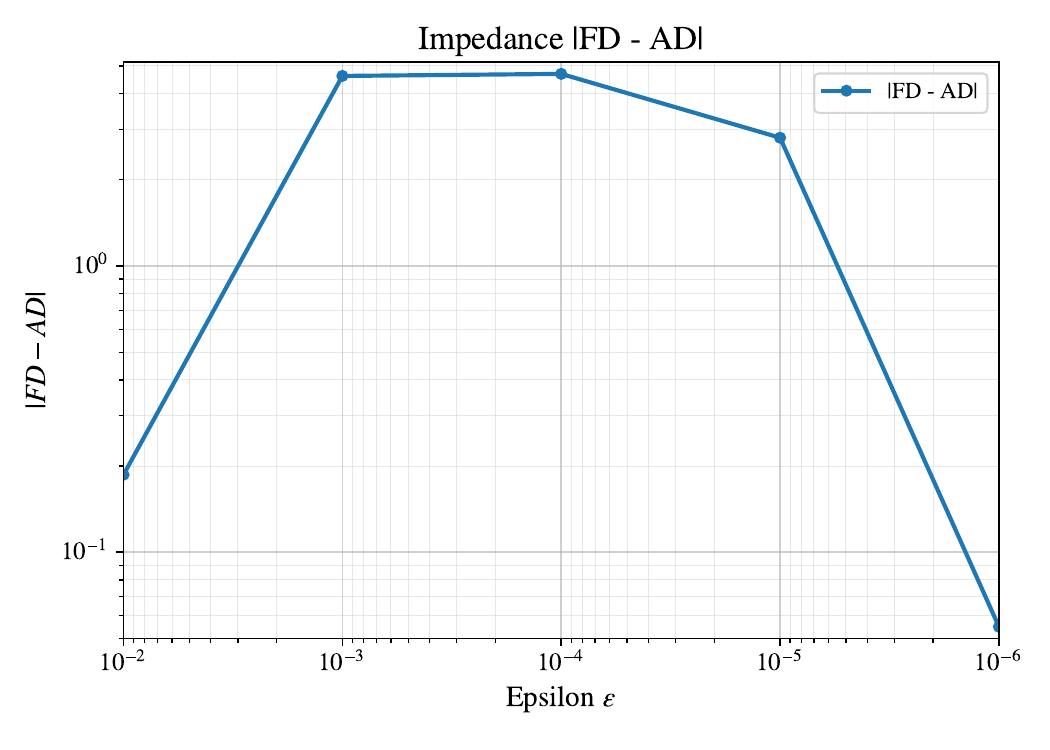}
        \caption{$Z$: $\lVert g_{\mathrm{FD}}-g_{\mathrm{AD}}\rVert$ vs. $\epsilon$.}
        \label{fig:grad_norm_Z}
    \end{subfigure}\hfill
    \begin{subfigure}[b]{0.33\linewidth}
        \centering
        \includegraphics[width=\linewidth]{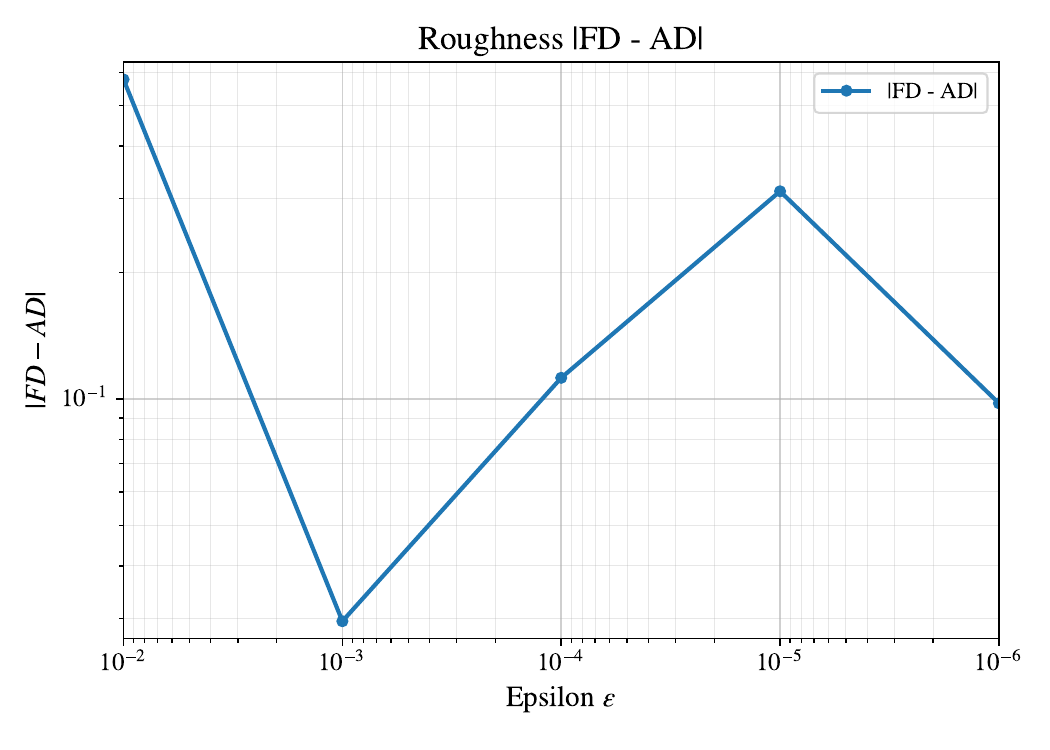}
        \caption{$\rho$: $\lVert g_{\mathrm{FD}}-g_{\mathrm{AD}}\rVert$ vs. $\epsilon$.}
        \label{fig:grad_norm_rho}
    \end{subfigure}\hfill
    \begin{subfigure}[b]{0.33\linewidth}
        \centering
        \includegraphics[width=\linewidth]{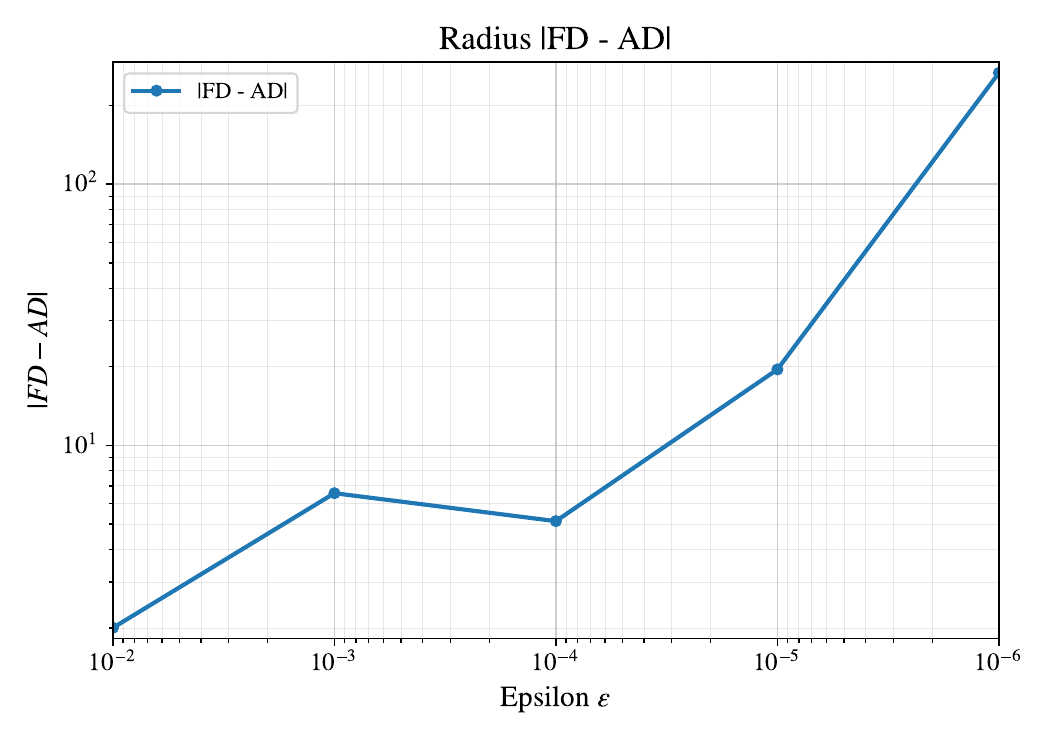}
        \caption{$R$: $\lVert g_{\mathrm{FD}}-g_{\mathrm{AD}}\rVert$ vs. $\epsilon$.}
        \label{fig:grad_norm_R}
    \end{subfigure}

    \caption{Automatic-differentiation (AD) and finite-difference (FD) gradient comparisons for the three optimized parameters in the simulation-to-real cylinder problem. The top row reports the AD and FD gradient values for impedance \(Z\), roughness \(\rho\), and radius \(R\) as functions of the FD step size \(\epsilon\), and the bottom row reports the corresponding discrepancy \(\lVert g_{\mathrm{FD}}-g_{\mathrm{AD}}\rVert\). The figure shows that FD agreement depends strongly on the perturbation size, with the closest correspondence occurring over an intermediate range of \(\epsilon\), thereby supporting the consistency of the AD sensitivities while also highlighting the limitations of FD validation at excessively small or large step sizes.}
    \label{fig:grad_compare_placeholder}
\end{figure}

From Figures~\ref{fig:real_param_histories} and~\ref{fig:real_loss}, the simulation-to-real optimization exhibits rapid early improvement followed by gradual stabilization. The weighted \(\ell_1+\ell_2\) loss remains near its initial level during the earliest iterations, then decreases sharply between approximately \(k\approx 25\) and \(k\approx 45\), indicating that the optimizer quickly enters a region of substantially improved agreement with the processed experimental target. After this transition, the loss continues to decrease overall, although with small oscillations, and approaches a near-plateau by roughly \(k\approx 120\)--\(145\). This behavior is consistent with practical convergence to a stable local optimum rather than continued large-scale parameter drift.

The parameter histories show that the three optimized quantities converge on different timescales. The impedance \(Z\) and roughness \(\rho\) evolve smoothly and monotonically during the early stage of the optimization before leveling off, suggesting that these parameters are well-informed by the image-space objective. In contrast, the radius \(R\) exhibits a more non-monotone trajectory, with an initial increase followed by a small overshoot and subsequent relaxation toward its final value. This indicates that the geometric parameter is more strongly coupled to the other scene parameters and may be more sensitive to local tradeoffs in the loss landscape. Taken together, these results suggest that the optimizer is able to identify a stable, effective parameter set that substantially improves agreement with the experimental target, while still reflecting the increased complexity and ambiguity of the simulation-to-real setting.

Figure~\ref{fig:grad_compare_placeholder} compares automatic-differentiation (AD) and finite-difference (FD) gradients for the three optimized parameters over a range of FD step sizes \(\epsilon\). An important feature of this comparison is that the AD pipeline is not constructed to differentiate through the stochastic Monte Carlo sampling process itself. Instead, the random numbers used to generate ray paths, scattering events, and receiver-sampling decisions are held fixed during the backward pass, so AD differentiates only the deterministic computation induced by a given sampled realization of the transport and image-formation pipeline. The corresponding FD comparisons are interpreted in the same conditional sense, namely as sensitivities with respect to the continuous parameters for a fixed Monte Carlo realization.

Overall, the results show that the agreement between AD and FD depends strongly on the chosen perturbation size, which is consistent with the expected tradeoff between truncation error at large \(\epsilon\) and numerical cancellation at very small \(\epsilon\). For the impedance gradient, the FD estimate changes sign and magnitude across the tested range, and the discrepancy \(\lVert g_{\mathrm{FD}}-g_{\mathrm{AD}}\rVert\) remains relatively large, indicating that the FD approximation is particularly sensitive to step-size selection for this parameter. The radius gradient shows comparatively better agreement at intermediate \(\epsilon\), but the discrepancy grows again for larger perturbations, suggesting that the local linear approximation deteriorates as the step size increases. The roughness gradient exhibits a similar trend, with the smallest mismatch occurring at an intermediate \(\epsilon\) and larger discrepancies appearing at both extremes of the tested range.

Taken together, these results support the consistency of the AD gradients in the simulation-to-real setting while also illustrating that FD-based validation is reliable only over a restricted range of perturbation sizes. More broadly, the figure highlights a central feature of the proposed differentiable formulation: the computed gradients represent sensitivities of the fixed-sample transport, beamforming, and post-processing pipeline with respect to continuous physical parameters, rather than derivatives of the underlying Monte Carlo randomness itself. This design avoids spurious gradient contributions from discrete sampling events and yields gradients that are substantially more stable and interpretable for optimization.

\begin{figure}[htbp]
    \centering
    \begin{subfigure}[b]{0.48\linewidth}
        \centering
        \includegraphics[width=\linewidth]{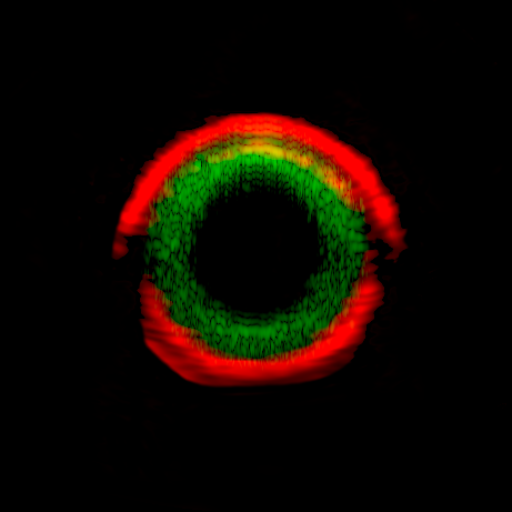}
        \caption{Initial comparison between the simulated and experimental images.}
        \label{fig:sim_real_initial_compare}
    \end{subfigure}\hfill
    \begin{subfigure}[b]{0.48\linewidth}
        \centering
        \includegraphics[width=\linewidth]{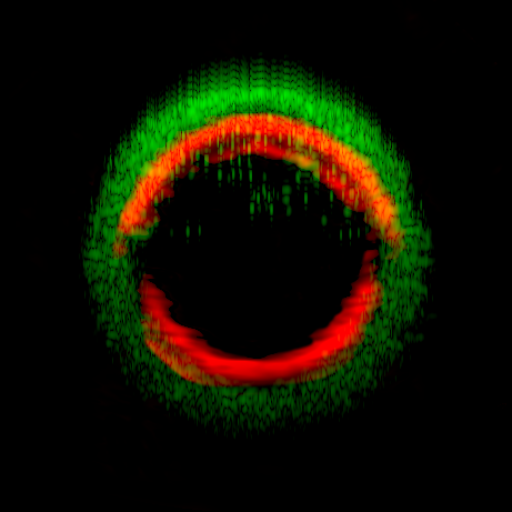}
        \caption{Final comparison between the optimized simulated and experimental images.}
        \label{fig:sim_real_final_compare}
    \end{subfigure}
    \caption{Overlay comparison of the simulated image (green) and the experimental target (red) before and after optimization. The initial comparison shows a clear mismatch in the location and extent of the dominant scattering boundary, whereas the optimized result exhibits improved alignment with the experimental ring response, particularly along the outer high-intensity interface that dominates the image-space loss.}
    \label{fig:sim_real_image_compare}
\end{figure}

Figure~\ref{fig:sim_real_image_compare} presents the initial and final simulated images overlaid with the experimental target used in the optimization. The optimized radius converges to \(R = 3.61\) mm, which is toward the upper end of the prescribed parameter range. This result is consistent with the geometric discrepancy between the simplified simulated model and the experimental target: the simulation represents an idealized zero-thickness interface, whereas the real ring has a finite physical thickness and therefore generates a broader reflected response in the measured B-mode image.

The observed behavior is further influenced by the logarithmic scaling inherent to B-mode imaging. Because the image-space objective is evaluated on a log-compressed image, the optimization is dominated by alignment of the most intense and visually prominent features. In this case, the strongest contribution to the loss reduction is obtained when the first major simulated reflection aligns with the outermost high-intensity boundary in the experimental image. Consequently, the optimizer favors a larger effective radius, yielding improved agreement in the dominant image structure even though the simulated geometry remains an idealized approximation of the physical ring.

\section{Conclusion}
\label{sec:conclusion}
In this work, we build on the existing UltraRay framework to develop a fully differentiable end-to-end ultrasound simulation pipeline based on full-path Monte Carlo ray tracing. The proposed formulation enables gradients to propagate from image-space objectives back through both the acoustic transport model and the downstream beamforming and post-processing stages, allowing optimization with respect to physically meaningful scene and acquisition parameters. In contrast to prior ray-based ultrasound simulators that have primarily been used as forward models, the present framework is designed explicitly for gradient-based calibration, inverse recovery, and acquisition optimization. A central feature of the approach is that, for fixed Monte Carlo samples, automatic differentiation is applied to the deterministic transport and image-formation pipeline with respect to continuous parameters, yielding stable and interpretable sensitivities for optimization.

The numerical examples demonstrate both the forward and inverse capabilities of the proposed framework. In controlled forward problems, the model reproduces expected geometric image features and captures more complex anatomical structures. In the simulated-reference setting, the framework successfully recovers known scene parameters from image-space supervision alone, confirming the consistency of the differentiable formulation. In the simulation-to-real setting, the method identifies an effective set of parameters that substantially improves agreement between simulated and experimentally acquired B-mode images, while appropriately reflecting the ambiguity and model mismatch inherent in real-data inversion. In addition, comparisons between automatic-differentiation and finite-difference gradients support that the computed sensitivities are meaningful for optimization and consistent with the underlying forward model.

It is important to note that, in the simulation-to-real setting, the recovered quantities should be interpreted as effective parameters rather than uniquely identified physical ground-truth properties. Because the experimental target only partially constrains the underlying scene and the simulation model remains intentionally simplified, multiple parameter combinations may produce similar image characteristics. The primary value of this experiment is therefore not the exact recovery of a unique physical parameter set, but rather the demonstration that the proposed framework can be meaningfully coupled to real experimental B-mode data and can identify simulation parameters that reproduce its dominant observed features.

At the same time, several limitations of the present work should be acknowledged. The current framework is based on a geometric ray-tracing approximation rather than a full-wave acoustic simulation and therefore does not capture the complete range of wave phenomena present in ultrasound imaging. In addition, the scene representation relies on simplified interface models, and the differentiable formulation computes gradients for fixed Monte Carlo realizations rather than differentiating through the underlying sampling randomness itself. A further limitation is that, in the simulation-to-real experiment, the simulated target is modeled as an idealized zero-thickness cylindrical interface, whereas the experimental cylinder has a finite wall thickness. This mismatch can broaden or shift the observed boundary response in the measured B-mode image and may cause the recovered radius, roughness, and impedance to act as effective fitting parameters rather than uniquely identifiable physical quantities. While these design choices are important for tractability and optimization stability, they also define the current scope of the method. Future work will focus on extending the framework to more complex tissue and interface models, including finite-thickness and volumetric targets, incorporating attenuation and volumetric scattering effects, enabling optimization over a broader set of acquisition parameters, and validating the approach on a wider range of experimental and anatomical imaging scenarios.

\bibliographystyle{unsrt}  
\bibliography{references}  






\end{document}